\documentclass[conference]{IEEEtran}
\IEEEoverridecommandlockouts

\usepackage[T1]{fontenc}

\usepackage[noadjust]{cite}
\usepackage{amsmath,amssymb,amsfonts}
\usepackage{graphicx} 
\ifCLASSOPTIONcompsoc
\usepackage[caption=false, font=normalsize, labelfont=sf, textfont=sf]{subfig}
\else
\usepackage[caption=false, font=footnotesize]{subfig}
\fi
\usepackage{bm}
\usepackage{textcomp}
\usepackage{xcolor}

\usepackage{algorithm,algpseudocode}

\usepackage{mathtools}    
\usepackage{makecell}  
\usepackage{amsthm}
\theoremstyle{definition}

\newtheorem{definition}{Definition}[section]

\RequirePackage{pgfplots}
\usepgfplotslibrary{groupplots}
\pgfplotsset{compat=1.9}

\graphicspath{"img"}

\title{Availability Analysis of Redundant and Replicated Cloud Services with Bayesian Networks}

\makeatletter
\newcommand{\linebreakand}{%
\end{@IEEEauthorhalign}
\hfill\mbox{}\par
\mbox{}\hfill\begin{@IEEEauthorhalign}
}
\makeatother

\author{
	\IEEEauthorblockN{Otto Bibartiu\hspace{0.5cm} Frank D{\"u}rr \hspace{0.5cm} Kurt Rothermel}
	\IEEEauthorblockA{
			\textit{University of Stuttgart} \\
			\textit{Institute for Parallel and Distributed Systems (IPVS)}\\
			Universit{\"a}tsstrasse 38
			Stuttgart, Germany \\
			otto.bibartiu@ipvs.uni-stuttgart.de \\
			frank.duerr@ipvs.uni-stuttgart.de \\
			kurt.rothermel@ipvs.uni-stuttgart.de}
	\and
	\IEEEauthorblockN{Beate Ottenw{\"a}lder \hspace{0.5cm} Andreas Grau}
			\IEEEauthorblockA{\textit{Robert Bosch GmbH} \\
				Stuttgart, Germany \\
				Beate.Ottenwaelder@de.bosch.com\\
				Andreas.Grau2@de.bosch.com}	
}

\begin{document}
	
	\maketitle

	\begin{abstract}
		Due to the growing complexity of modern data centers,  failures are not uncommon any more. Therefore, fault tolerance mechanisms play a vital role in fulfilling the availability requirements. 
		Multiple availability models have been proposed to assess compute systems, among which Bayesian network models have gained  popularity in industry and research due to its powerful modeling formalism. 
		In particular, this work focuses on assessing the availability of redundant and replicated cloud computing services with Bayesian networks. 
		So far, research on availability has only focused on modeling either infrastructure or communication failures in Bayesian networks, but have not considered both simultaneously.
		 This work addresses  practical modeling challenges of assessing the availability of large-scale redundant and replicated services with Bayesian networks, including cascading and common-cause failures from the surrounding infrastructure and communication network. 
		 In order to ease the modeling task, this paper introduces a high-level modeling formalism to build such a Bayesian network automatically.
		 Performance evaluations demonstrate the feasibility of the presented Bayesian network approach to assess the availability of large-scale redundant and replicated services. 
		This model is not only applicable in the domain of cloud computing it can also be applied for general cases of local and geo-distributed systems.
	\end{abstract}
	

\section{Introduction}

Due to the growing complexity of modern data centers, failures are not the exception anymore; they are the norm~\cite{cotroneo19}. For example, the OVHcloud data center incident in 2021 led to the unavailability of multiple online businesses~\cite{ocloud21}, while the Facebook outage in late 2021, caused by a miss-configuration of the backbone routers~\cite{janardhan21}, led to an estimated loss of 65 million dollars in revenue~\cite{brown21}. 
Cloud operation teams and reliability engineers employ fault tolerance techniques to mask faults through redundancy or replication, by deploying multiple instances of the same service to increase availability.
These instances are not fault independent. They normally share common cause failures with the surrounding execution environment and communication network, raising the question  if  fault tolerance measures meet the availability requirements. To answer this question, this paper proposes a novel Bayesian network modeling appraoch to assess the availability of redundant and replicated cloud services  in presence of network and common-cause failures.

This work distinguishes between the terms redundant and replicated cloud services to address two different modeling semantics with respect to service communication, which can lead to different availability outcomes.
In a broader sense, redundancy implies independent service instances (copies) that work in parallel. Redundant services can be stateful or stateless. For example redundant DNS servers are stateful, where multiple DNS  instances can independently serve client requests. Stateless redundant services are AWS Lambda and Azure Functions, which are part 
of the Function-as-a-Service (FaaS) layer. 
In contrast, replication always involves stateful services that implement a replication protocol to maintain the desired degree of state consistency between the instances. 
Examples of such systems are replicated databases~\cite{lakshman10,pstore,alpos20}, and distributed locking services~\cite{burrows06}. 
These instances need to communicated with each other at some point in time as supposed to instances of a redundant service. 

\begin{figure} 
	\centering
	\subfloat[Redundant system represented by an independent set of instances which all ofer the same service.\label{1a}]{%
		\includegraphics[width=0.7\linewidth]{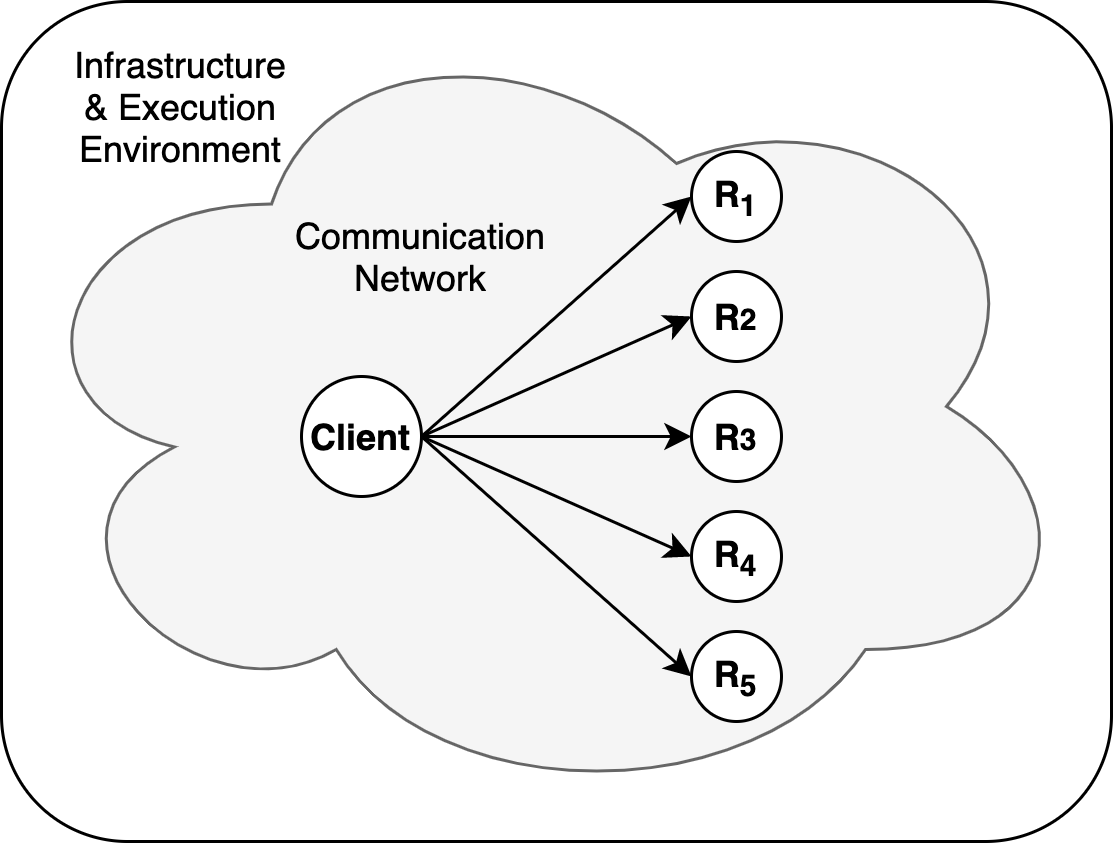}}
	\\
	\subfloat[Stateful replication which necesitates communication between replicas to aggree upon the state of the service.\label{1b}]{%
		\includegraphics[width=0.7\linewidth]{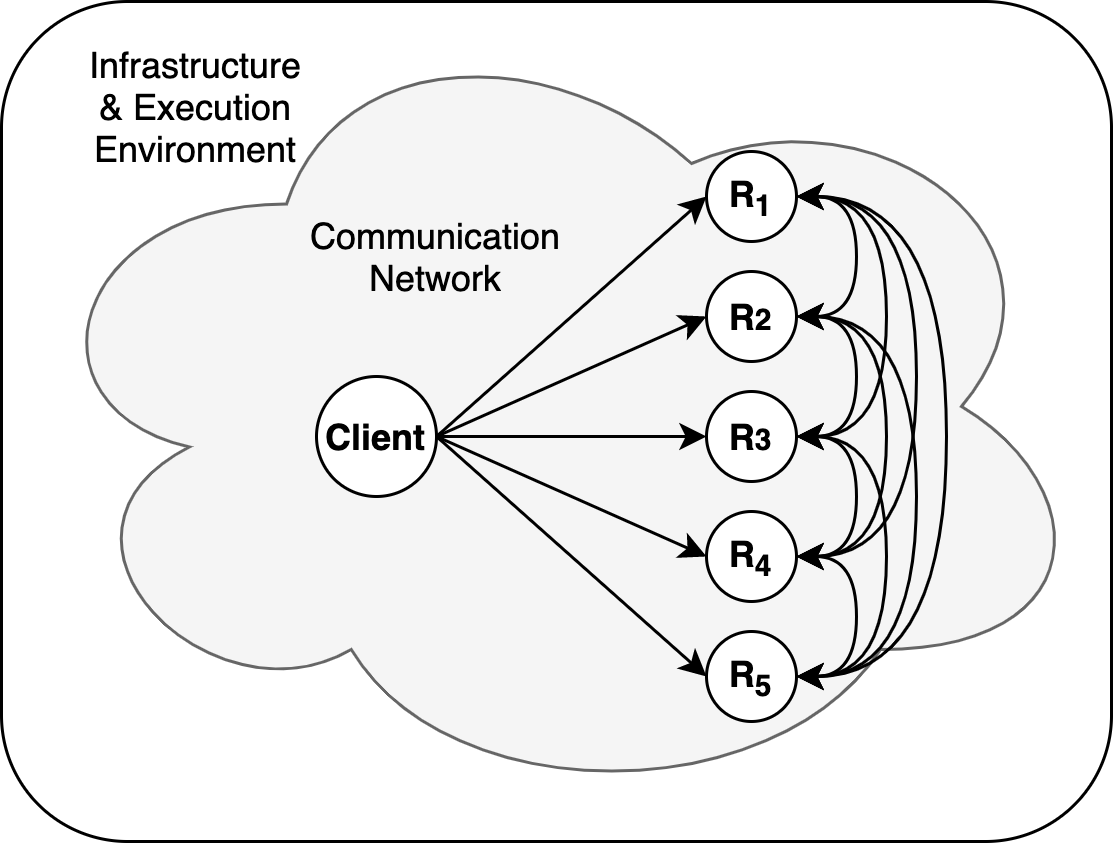}}
	\caption{To assess the availability of a redundant or replicated service, one needs to consider the reachability of the  instances through the communication network, as well as their fault dependencies with the execution context}
	\label{fig:system} 
\end{figure}

The ISO/IEC/IEEE International Standard on Systems and Software Engineering defines availability as the ``degree to which a system or component
is operational and accessible when require''\cite{ieee17}. 
Similarly, we refer to availability as the likelihood of a cloud service to be reachable and operational (up) when required.   
Figure~\ref{fig:system} exemplifies the difference in failure modes when assessing the availability of a redundant or replicated services.
Common cause failures and cascading faults in the infrastructure can simultaneously lead to the unavailability of multiple service instances. 
Network faults might lead to network partitioning, which renders services instances unreachable for client requests or segmenting the instances of a replicated service into groups that cannot agree upon the next states.
For example,  Figure~\ref{1a} shows a redundant services. A client application regards the redundant service  available as long as it can reach at least one of the instances. 
In contrast, Figure~\ref{1b} depicts a replicated service, which has the overhead of inner-replica communication due the necessity of implementing a replication protocol. 
So, the replicated service is reachable as long as at least one working instance is reachable by the client, and the instance can communicate with sufficient remaining instances to reach the required quorum size, i.e. to correctly implement the replication protocol. 
As a result, this communication overhead might involve more network components that form an  additional source for potential failures, which we need to account for in our availability model.  

As Michael R. Lyu noted~\cite{lyu07}, it is not sufficient to assess the reliability or availability of a software system in isolation. It is important to also consider the execution (operational) environment, in order to create accurate availability models. 
However, while researchers acknowledge the significance of  infrastructure and communication faults~\cite{garraghan18,gunawi14}, they usually model either the infrastructure~\cite{kim11,jammal16,bennacer15}, or the
communication~\cite{ghosh14,epstein16,ford10,chiang21} part of a system. 
Moreover, with the advent of cloud computing, reliability engineers face the challenge of modeling the availability of large-scale cloud services. Especially with the introduction of FaaS in cloud computing and NoSQL databases, such as Cassandra, the number of instances per service has grown in the hundreds~\cite{netflix}. Consequently, a high number of components lead to an increase in structural complexity, making many availability models impractical or render them infeasible to model large-scale cloud services. 

Consequently, in order assess the availability of today's cloud services, we need  holistic  availability models that can model large-scale replicated and redundant cloud services while simultaneously accounting for cascading and common-cause failures of the network and infrastructure environment.
This paper addresses this problem by proposing a Bayesian network availability model. 
Bayesian networks  have proven helpful in computing the availability of complex systems since they provide a powerful modeling formalism to express complex fault dependencies and uncertainty between components~\cite{langseth07,duan12,pan17}. 
They support a rich set of efficient inference algorithms suitable for fault diagnostic~\cite{bennacer15} and availability prediction. Moreover, with the introduction of scalable Bayesian network structures~\cite{bibartiu21,heckerman93}, we argue that Bayesian networks are  a good fit to assess large-scale redundant and replicated cloud services.

This work provides the following contributions.
\begin{enumerate}
	\item We introduce a high-level modeling formalism to describe complex redundant and replicated cloud services at any preferred level of infrastructure and network granularity, since manually building a  Bayesian network availability model of large-scale services can become tiresome, time-consuming, and error-prone (this model gets then translated into the Bayesian network model later). 
	\item We explain step-by-step how to address the modeling challenges of implementing a Bayesian network model that considers cascading infrastructure and network communication failures.  
	\item Especially for replicated services, we solve the modeling challenge of addressing network partitioning failures, while also considering a flexible range of fault tolerance semantics like voting and weighted-voting based replication. 
	\item We also propose a translation procedure that transforms the high-level model into the proposed Bayesian network availability model automatically. 
	\item  Finally, we provide evaluations that demonstrate the feasibility of building and assessing large-scale cloud services models with hundreds of infrastructure components and service instances.  
\end{enumerate}

The remainder of this paper is structured as follows:
In Section~\ref{sec:systemmodel}, we introduce our system assumptions. 
Next, in Section~\ref{sec:model},  we formulate our high level availability model.
Afterward, in Section~\ref{sec:bn}, we show how to build the Bayesian network available model.
In Section~\ref{sec:eval}, we evaluate the performance of our Bayesian network approach to model large scale services.
Next, in Section~\ref{sec:discussion}, we discuss the results and suggest future work topics.
In Section~\ref{sec:rw}, we present related work on  availability modeling of replicated systems.
Finally, in Section~\ref{sec:conclusion}, we conclude this paper.

\section{System Model}
\label{sec:systemmodel}

The proposed availability model considers redundant or replicated distributed (cloud service) systems as a set of instances. Instances are assumed to run on virtual or physical hosts, placed within the infrastructure of one or more data centers, and linked by a communication network. The network is assumed to consist of components such as switches, routers, and middleboxes, e.g., firewalls, which are placed within the same infrastructure as the hosts themselves. 

Specifically, redundant services can be  stateless or stateful services, where the stateful service does not replicate its state. Replicated services always refer to stateful services where state is replicated.

A replicated service is available when sufficient replicas are available. Conversely, if too many replicas are unavailable, i.e., have crashed or are not reachable, the service is considered unavailable at the time of the request.  
A quorum is a certain set of k-out-of-n redundant instances that need to be available to provide a particular service function. Note that different functions such as reading or writing a data object can have different quorum sizes,  depending on the replication protocol. Therefore, in this work, service availability  implicitly refers to the availability of a specific service function or operation.  

The model considers two types of communication patterns. For redundant services, we assume that a client only needs to communicate with one instance to issue its request.
For replicated services, it is also sufficient for a client to communicate with one instance to initiate the request. However, that instance needs to be able to communicate with sufficient  remaining instances to agree upon the result of the client's request. 
The exact fault tolerance semantics for redundant and replicated services  is flexible and can be defined by the reliability engineer  as part of the system description.

The hosts and the communication network are part of the infrastructure, which forms a complex component-based system consisting of  \textit{infrastructure components}, such as data centers, racks, power supplies, virtual machines, and network appliances. The model assumes that hard -- and software -- components, including the service instances, have a crash-recovery model. As soon a component encounters a failure, it crashes and stops, and recovers eventually. Each component in the infrastructure has its probability of failing by its own without external influence. 

Moreover, the model assumes that infrastructure components have fault relations, representing potential common causes of failures. These fault relationships can form a cause-effect chain, where the failure of one component is the cause of failure of another component, essentially propagating the failure through the infrastructure, until it eventually leads to the failure of the cloud service, i.e., cascading failure. 
 In order to formalize the  relationship between two directly fault dependent components, the model assumes that the dependence can be described by means of a static fault tree~\cite{ruijters15}. 

Client applications and instances can communicate with each other by exchanging messages via the communication network. The network is composed of network components forming a network graph. The end-to-end communication, i.e., channels, between instances and clients can be synchronous or asynchronous and implemented by one or more redundant network routes. A channel crashes when there is no route in the network to connect the two endpoints, and a route becomes unavailable when at least one network component along the route crashes. Client applications might be placed outside of the known infrastructure. In this case, the model considers the paths starting from the network appliance that constitutes the entry point of the data center; or, if the client application is within the data center, its host.
Moreover, we assume the exists some dedicated network components, e.g., firewalls or load balancers, that act as  \textit{gateways}, i.e. entry points, for clients applications to communicate with the service. 

A particular placement of instances to virtual or physical hosts is called a \textit{deployment} and known beforehand. Instances do not migrate. If an instance crashes, it does not recover on a different host. It recovers back at its former host. Hence, if a host crashes, all its instances can recover when the host recovers.
The model makes no restrictions on the number of instances per host. Multiple instances can run on the same host. 
In the case of replication, the model does not assume the concurrency control method or the particular replication protocol. Either at any given point in time there are enough replicas \textit{up} and \textit{reachable} to agree upon the results of a client's request, or too many replicas crashed or are unreachable, such that the remaining replicas cannot form a quorum for any client request, resulting into the unavailability of the service.

\section{High Level Model Description}
\label{sec:model}
This section will address the modeling challenge of building a Bayesian network model to infer the  availability of a cloud service in the presence of cascading infrastructure and network faults. 
To ease the modeling process, we present a high-level model description first, which we later translate to a Bayesian network. 
The model contains three basic sub-models: a failure model for the infrastructure, a model for the network, and a model to describe the fault-tolerance semantics of the service. 
This provides the advantage to choose the component granularity of the system.
First, we begin with the basic unit of our model, a component.
\begin{definition}[Component]
	A component $C \in \bm{C}$, from the finite set of all  components of the system $ \bm{C} = \{C_1, C_2,\dots\}$, is an indivisible hard or software entity with the states $\{F,T\}$, and a probability distribution  $P(C = F) = q_i$ to observe the component as faulty (unavailable) and $P(C = T) = 1- q_i$ to observe  the component as operational or working (up). 
\end{definition}
The set $\bm{I} = \{ I_1, \dots,I_n \} \subset \bm{C}$ are instances of the service. The remaining components are infrastructure and network components. 

Components might have fault dependencies between themselves. We describe these fault dependencies as a direct acyclic graph (DAG).
\begin{definition}[Fault Dependency Graph]
	Given the set  of all components  $ \bm{C}$, the model defines the fault dependency graph as
	a DAG $G_{\text{FD}} = (\bm{C},E_{\text{INF}},FT)$, with edges $E_{\text{INF}} \subseteq  \bm{C} \times \bm{C}$, and an associated (static) fault tree model $FT$ for every component in $\bm{C}$.
\end{definition}
Directed edges are tuples $(C_i,C_j)$, where $C_i$ is said to be a parent component of $C_j$, and $C_j$ is said to be a child component of $C_i$.   These edges can also define a \textit{contained-in} relation, to signify that one component is contained within another.

\begin{figure}[ht]
		\centering
		\subfloat[Fault dependencies of a host. \label{fig:ftexample:graph}]{%
		\includegraphics[scale=0.50]{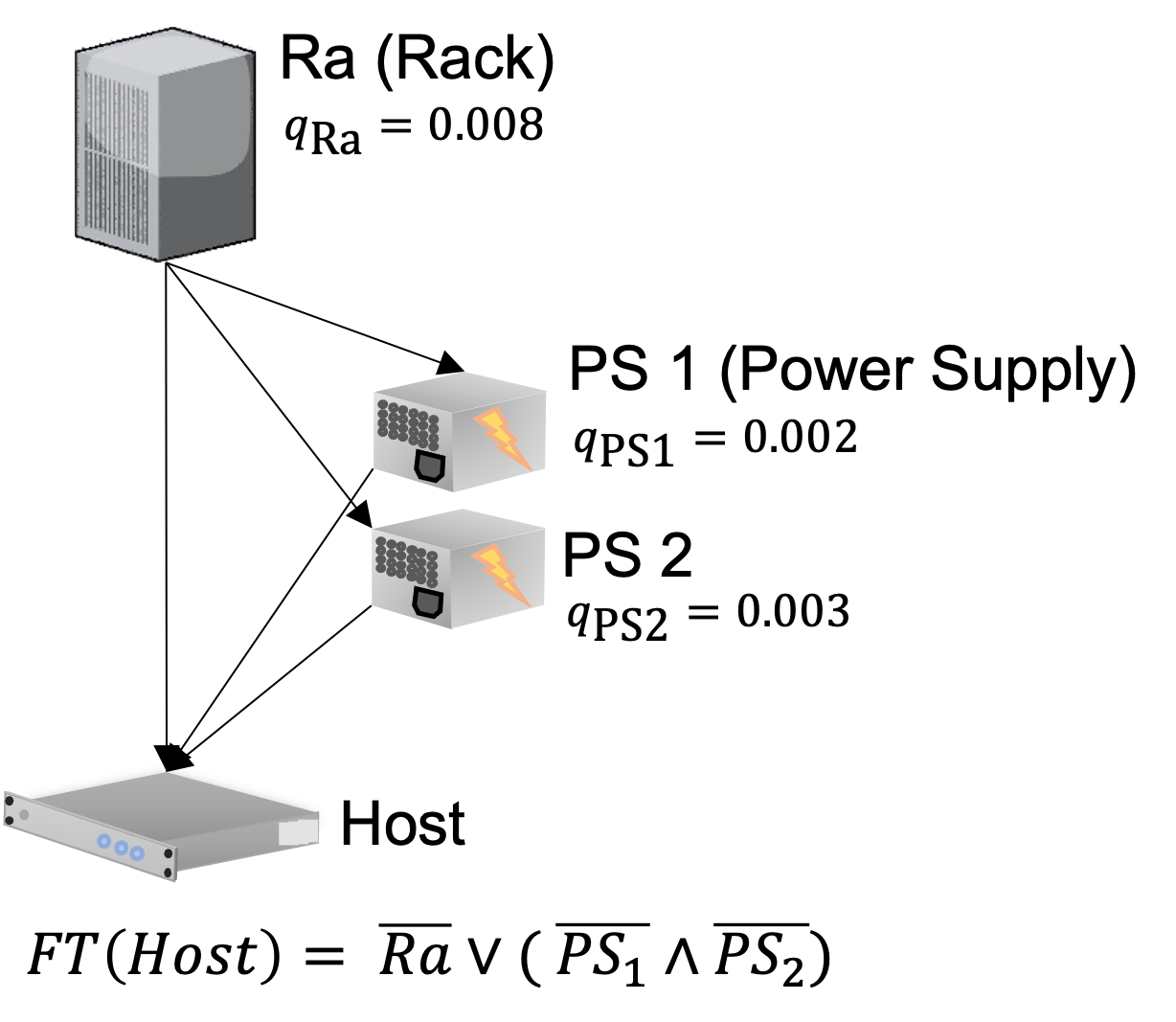}}
	\hfill
\subfloat[Local fault tree model of the host component. \label{fig:ftexample:ft}]{%
		\includegraphics[scale=0.10]{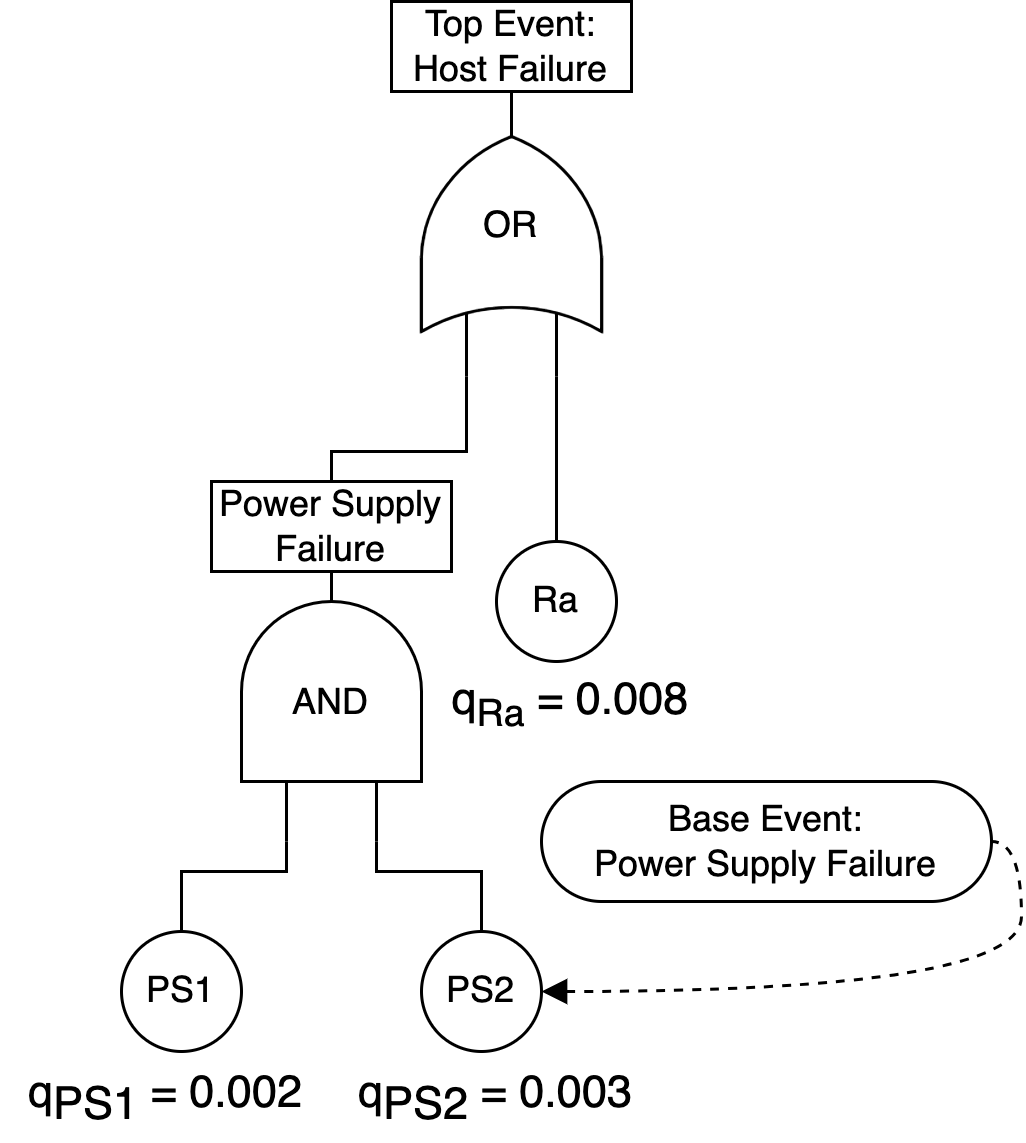} }
	\caption[Fault Dependency Graph Example]{Fault Dependency Graph Example.}
	\label{fig:ftexample}
\end{figure}

In order to express complex component dependencies, $FT(C_i)$ contains the definition of a static fault tree that describes the fault semantics of a component $C_i$ as a function of its parent components.  $FT(C_i)$ has as the top event the failure of component $C_i$ and as base events $C_i$'s parents components.
For example, Figure~\ref{fig:ftexample:graph} shows a fault dependency graph with a host component that depends on three parent components to illustrate how to apply $FT$. 
The host fails if the rack breaks, e.g., catching fire, or both power supplies stop working. $FT(\text{host})$  encodes this failure relation at the host component, as shown in Figure~\ref{fig:ftexample:graph}. Hence, the corresponding fault tree shown in  Figure~\ref{fig:ftexample:ft} has the power supplies and the rack as basic input events and the host failure as the top event. The fault tree uses an OR gate to trigger the top event. The hosts fails when the rack fails, or both power supplies fail, represented by the AND gate. The fault tree is part of the host component to determine the cause of a host failure due to  external influences, which depends solely on its parent components. 
Note that the fault dependency model is a DAG, disallowing cyclic fault dependencies since it leads to cycles in the final Bayesian network graph, which is not allowed by definition. 

To account for communication faults, the model needs a  representation of the network.
Network components represent network appliances such as switches, routers, load-balancers, and firewalls. Consequently, the failure of related infrastructure components can influence the failure of a network component, which can lead to communication failures. Unlike the fault dependency graph, the network graph can have cycles. 

\begin{definition}[Network Graph]
	Given a set of hosts $H \subset \bm{C}$,  a set of network components $N \subset \bm{C}$, and their union $\bm{C}_{\text{NET}} = H \cup N$, the network is a graph  $G_{\text{NET}}= (\bm{C}_{\text{NET}},E_{\text{NET}}) $ with unidirectional edges, where the edges $E_{\text{NET}} \subseteq \bm{C}_{\text{NET}} \times \bm{C}_{\text{NET}}$ define the communication links between any two network components.
\end{definition}

With this graph notion, reliability engineers can decide the granularity of the network model. Suppose they have little or no knowledge of the network. In that case, they can represent the network as 'one switch' connecting all instances, aggregating all potential failure probabilities as one value for one \textit{super} component. However,  they can also describe more complex network graphs if they have ample knowledge, which improves the model w.r.t. a more realistic representation of the actual network. 

The final system description of the cloud service  is the unification of the above model definitions. 
\begin{definition}[High-level System Model]\label{def:S}
	A system
	$$	S = (\bm{C},Q,G_{\text{FD}},G_{\text{NET}},D,P,\mathcal{G},c) $$
	is a eight-tuple consisting of the following elements:
	\begin{description}[\IEEEsetlabelwidth{Service}\IEEEusemathlabelsep]
		\item[$\bm{C}$] The set of all infrastructure, network components ans instances.
		\item[$Q$] The fault tolerance model defined as a path set of instances $Q = \{Q_1, \dots, Q_m\} \subseteq 2^{\bm{I}}$.
		\item[$G_{\text{FD}}$] The fault dependency graph.
		\item[$G_{\text{NET}}$] The network graph.
		\item[$D$] The association of instances to hosts $D: \bm{I} \rightarrow H$. 
		\item[$P$] The function of all fault probabilities of the components in $\bm{C}$.   
		\item[$\mathcal{G}$] The set of network components that  act as entry point for client applications to establish a communication channel with the instances of the services. $\mathcal{G} \subseteq C_{\text{NET}}$.
		\item[$c$] A Boolean value $c \in \{\textit{false,true}\}$ to indicate if the service is redundant or replicated.
	\end{description}
\end{definition}

The parameter $Q$ defines all instance combinations for which the service is considered in a working state in the presence of instance failures. This generic definition fits redundant as well as replicated service.
It  implies the enumeration of all valid instance combinations to build $Q$, building  a (minimal) path set of the service instances.  For example, let us assume a service has three instances $\bm{I} = {I_1,I_2,I_3}$ and the service works as long as two instances are up. As a result, $Q$ is the enumeration of all combinations with at least two instances $Q = \{\{I_1, I_2\}, \{I_1, I_3\}, \{I_2, I_3\}, \{I_1, I_2, I_3\}\}$. This definition provides a flexible way to express a wide range of fault tolerance semantics.
However, the enumeration of all instance combinations can become inefficient, especially when considering services with hundreds of instances. To alleviate this burden, we suggest an implicit construction method for k-out-of-n redundancy and voting-based replication models, as well as for the special cases of read-one and write-all replication. For these specific models, we define $Q$ as a tuple $(V,t)$, where $V = (v_1,...,v_n)$ are instance votes and $t$ a threshold value. The availability model will then account for the probability of observing sufficient working instances such that their votes exceed the threshold. For example, we can express the previous examples as $Q = ((1, 1, 1), 2)$ to implement the majority set without enumerating all possible set combinations.
If the service has different thresholds, that is, different quorum size requirements, per operation like read-one write-all replication.  Read-one would have $t = 1$ for the read operation and write-all $ t = n$ for the write operation. The service definition would then refer to one specific operation. Multiple operations can be supported by defining a service model for each operation separately and compute their availability values. At this point, it is up to the reliability engineers how to aggregate the availability of the different operations. They can use the lowest resulting value as a means to assess the probability of the worst-case service model, or they could compute the (weighted) average availability across all operations. Independently of what aggregation method a they chooses, this work shows how to build the availability model accordingly.

\begin{figure}[h]
	\centering
	\includegraphics[scale=0.35]{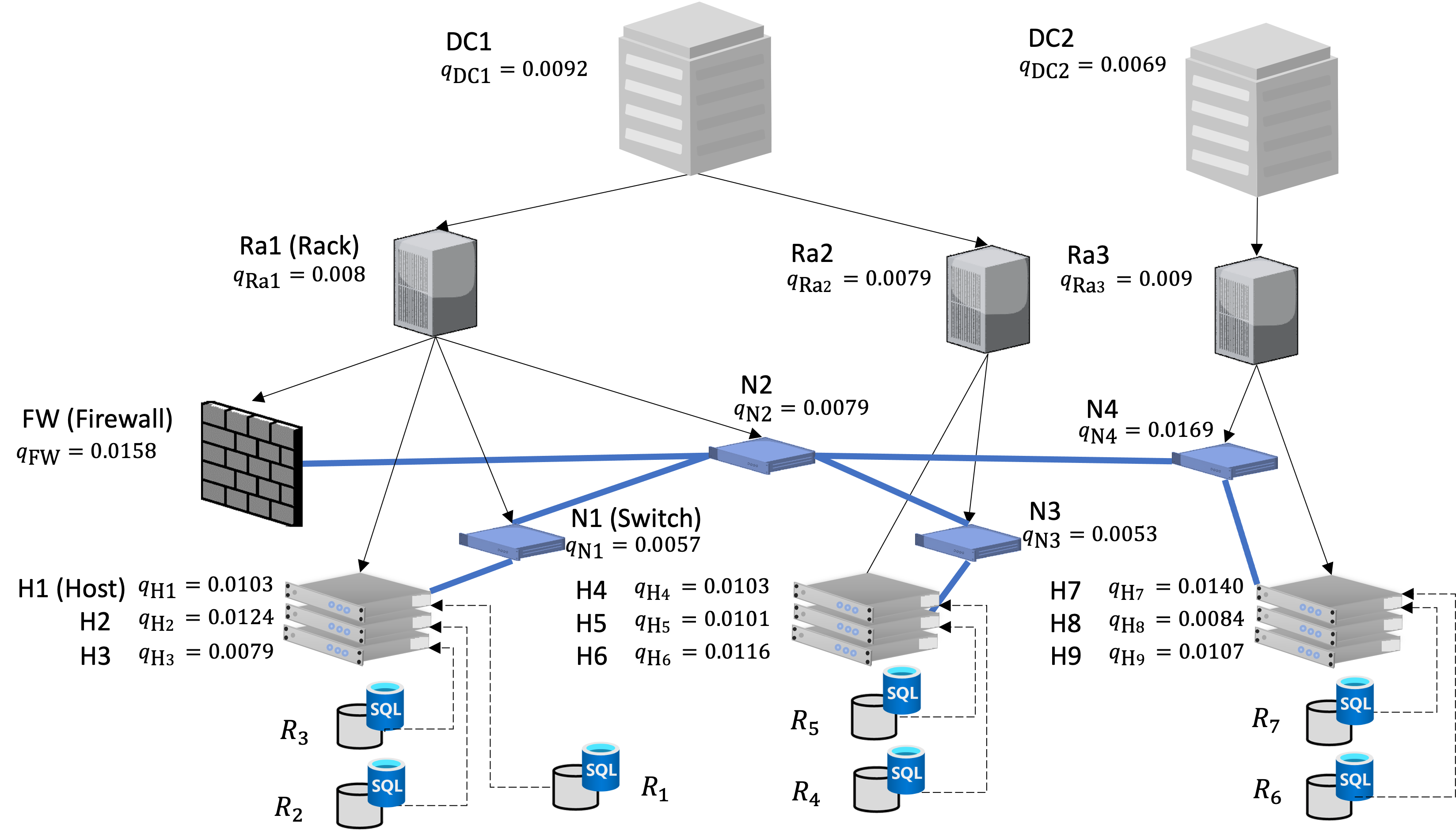}
	\caption{ Database management system example.}
	\label{fig:infrastructure}
\end{figure}

Let us exemplify the system model by describing a database management system as shown in  Figure~\ref{fig:infrastructure}, which
we will then use as a running example for the construction of the Bayesian network model next section.

Figure~\ref{fig:infrastructure} shows the overall system with its infrastructure and network components that provides the execution environment for the database management system. Although the data center infrastructure might be much larger, we only consider those components which serve the service. 
The database management system consists of seven replicas $ I_1$ to $I_7$, placed on hosts within the infrastructure of two data centers. Without loss of generality, the service is available as long as the replicas can form a majority quorum. 

Black arrows define fault dependencies between infrastructure components and blue edges represent communication links between network components. Without restrictions, in this example, we assume that a component fails when all its parent components fail; however, our  Bayesian network model will also be capable of modeling more complex component dependencies, such as redundant power supplies. Each component has its own intrinsic fault probability $q$ representing the likelihood of a component failure without external influence. Here, the fault probabilities are sampled from a beta distribution with $\forall i: q_i \sim \mathcal{B}(10, 1000)$. 

Finally, the  database management system  has the following  service description:
$$S_{\text{Example}} = (\bm{C},Q,G_{\text{FD}},G_{\text{NET}},D,P,\mathcal{G},c)$$
\begin{itemize}
	\item The set of all components is 
	\begin{equation*}
		\begin{split}
			\bm{C} = &\{DC_1, DC_2,
			Ra_1, Ra_2, Ra_3, 
			FW, N_1, N_2, N_3, N_4, \\
			& H_1, H_2, H_3, 
			H_4, H_5, H_6, \\
			& H_7, H_8, H_9,
			I_1,I_2,I_3,I_4,I_5,I_6,I_7\}
		\end{split}
	\end{equation*}
	
	\item  For the majority set, we need to form all combinations of at least four replicas.  
	$Q = \{  \{ I_1,I_2,I_3,I_4 \}, \{ I_2,I_3,I_4,I_5 \}, \dots \}$. 
	Or we can use the short hand notation $Q = ((1, 1, 1,1,1,1,1), 4)$.
	\item The deployment of replicas to hosts is given by the function $D$.
	\begin{eqnarray*}
		D(I_1) = H_1 \quad D(I_2) = H_2  \quad D(I_3) = H_3  \\
		D(I_4) = H_4 \quad  D(I_5) = H_5  \quad D(I_6) = H_7 \\
		D(I_7) = H_7 
	\end{eqnarray*}
	
	\item The fault dependency graph has the following definition. 
	$$G_{\text{FD}} = (	\bm{C},E_{\text{INF}},FT)$$ 
	\begin{eqnarray*}
		E_{\text{INF}} =& \{ (DC_1, Ra_1), (DC_1, Ra_2) , (DC_2, Ra_3),  \\ 
		& (Ra_1, FW), \dots , ( D(I_4), I_4), \\ 
		& (D(I_5),I_5), (D(I_6),I_6), ( D(I_7),I_7)	\}	
	\end{eqnarray*}
	Here, the instances use the deployment function to identify their host within the fault dependency graph.
	
	In this example, a component automatically fails when its parent component fails. So, $FT$ is a simple mapping of the failure event of the parent component of $C$, denoted as $pa(C)$. 
	$$ \forall C \in \bm{C}: FT(C) = \bigwedge_{C_i \in pa(C)}( C_i = F)$$ If a component has no parent, e.g. $DC_1$, then $pa$ returns the empty set. 
	
	\item  The network graph has the following form.
	$$G_{\text{NET}} = (C_{\text{NET}},E_{\text{NET}})$$
	\begin{eqnarray*}
		C_{\text{NET}} =	&\{FW, N_1, N_2, N_3, N_4, 
		H_1, H_2, H_3, 
		H_4, H_5, \\
		& H_6, H_7, H_8, H_9\}
	\end{eqnarray*}
	\begin{eqnarray*}
		E_{\text{NET}} = &	\{	\{FW,N_2\},	\{N_2,  N_1\},\{N_2,  N_3\}, \dots,\\
		&\{N_4,	H_7\},	\{N_4,H_8\},	\{N_4,H_9\}\}
	\end{eqnarray*}
	
	\item The fault probabilities of observing the  components as unavailable are:   
	\begin{eqnarray*}
		P(DC_1 = F)= 0.0092 \quad P(DC_2 = F)= 0.0069   \dots \\
		\quad P(H_8 = F)= 0.0084   \quad P(H_9 = F)=  0.0107\quad\quad
	\end{eqnarray*}
	For the sake of readability, we assume that instances do not fail due to intrinsic faults. Hence,
	they have an availability of one.
	
	\item The entry point for client applications is the firewall: $\mathcal{G}= \{FW\}$
	
	\item With $c=\textit{true}$, the model will consider communication between the instances, describing a replicated service. 
\end{itemize}

For example, the final model would address failure modes where rack $Ra_1$ would fail, which leads to the failure of all its built-in components. This includes its hosts $H_1$ to $H_3$, the firewall, and the switches $N_1$ and $N_2$ to fail as well. As a result, the replicas $I_1$ to $I_3$ would also fail since $Ra_1$ is a common cause of failure here. 
The Bayesian network model compactly encodes all combinations  of component state and their probabilities, for which the service is considered available,  as part of its qualitative representation, without enumerating all potential failure combinations explicitly.

\section{Bayesian Network Model}
\label{sec:bn}
The translation of the high-level service model into a Bayesian network consists of three steps. 
First, it builds a Bayesian network model of the fault dependency graph. 
Afterward, it extends the initial Bayesian network with the failure model for inner-replica communication when considering replicated services.
The third step finalizes the Bayesian network model by including the failure model for the client-to-instance communication. 
This modeling approach is novel insofar it can address network partitioning failures, which defines the availability of the service as a function of the channels between instances.
For instance, in the case of replicated services with voting-based replication, instead of building a model that accounts for at least k-out-of-n working instances, we build a model where we can infer the probability that for any reachable instance, there are at least (k-1)-out-of-(n-1) working channels connected to the remaining working instances. 

\subsection{Background}
We will use the Bayesian network representation of fault tree gates throughout the modeling process. This section provides the necessary background to understand fault trees and their equivalent Bayesian network notation. Readers familiar with this notation are free to skip this subsection. 

Fault trees are graphs that describe how certain combinations of component faults, known as base events, can lead to an undesired system failure, known as the top event. Logic gates are used to create intermediate events by forming a Boolean expression to describe what combinations of base events lead to a system failure ~\cite{stamatelatos02}. 
There are three basic gate types that have all fault tree variants in common: the AND, OR, and the k-out-of-n voting gate. The AND gate propagates a fault if all input events trigger a fault, while the OR gate propagates a fault when one input event has triggered a fault. The k-out-of-n voting gate propagates a fault when more than k-out-of-n inputs are fault events. 
The voting gate is suitable to model groups of redundant components. A group is considered available as long as no more than $n-k+1$ components are available, with $k$ being the number of failed components.

Bobbio et al.~\cite{bobbio01} introduced the general approach to represent fault tree gates with the help of Bayesian networks. This work will use the translation concepts as templates to construct the proposed Bayesian network availability model.  
A discrete Bayesian network~\cite{pearl88} is a DAG $G = (X,E)$ that represents a joint probability distribution  $P(X)$ over the set of discrete  random variables  $X = \{ X_1, X_2, \dots X_n\}$.
The term \emph{variable} or \emph{node} are used interchangeably to denote the vertices of the Bayesian network graph. 
For every edge $(X_i,X_j) \in E$ between the nodes $X_i$ and $X_j$,  $X_i$ is said to be a parent node of $X_j$, and $X_j$ is a child node of $X_i$. Each variable has a conditional probability distribution  $P(X_i=x_i | \text{pa}(X_i))$ encoded as a conditional probability table (CPT). The CPT contains the probability to observe a certain state $X_i=x_i$ given the observed states of its parent nodes denoted by parent function $\text{pa}(X_i) = \{X_p: \forall  (X_p,X_i) \in E \}$. Nodes without parents are called root nodes and have an a prior probability distribution $P(X_i=x_i)$. 

A Bayesian network entails a full joint probability distribution compactly as the product of all the nodes' conditional probability distributions:
\begin{equation}
	P(X) = \prod_{x \in X} P(x| \text{pa}(x))
	\label{eq:jp}
\end{equation}
With the help of the joint probability distribution, one can use inference to compute the posterior distribution $P(Y| X')$ of some query $Y \subset X$  of uncertain variables from a  given subset $X' \subset X\backslash Y$  of observations of the remaining variables.

\begin{figure}[h]
	\centering
	\includegraphics[width=\linewidth]{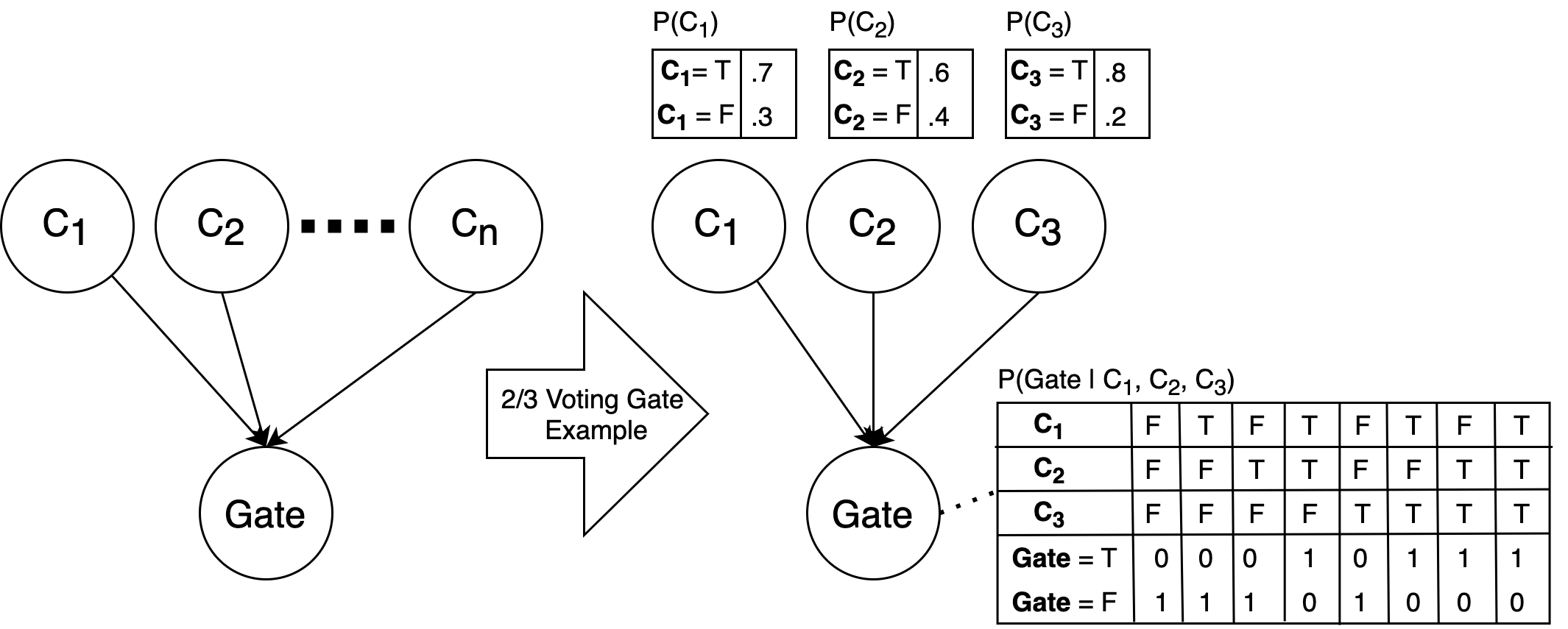}
	\caption[Basic Bayesian network to represent fault tree gates]{Basic Bayesian network to represent the fault tree's AND/OR, or k-out-of-n voting gates (left). Example instance of a Bayesian network k-out-of-n model (right).}
	\label{fig:simplebn}
\end{figure}

Figure~\ref{fig:simplebn} (left side) shows the main Bayesian network structure to realize the AND/OR and the k-out-of-n voting gate.
The basic structure has  $n$ components $C_1$ to $C_n$ with prior probabilities represented by their eponymous binary random variables with states $\{F,T\}$,  observing the component either faulty or available, respectively.  
The individual semantics of the gate types are encoded within the CPT of the $\operatorname{Gate}$ node.
The fault relation of fault tree gates is defined over the fault state of their input events, making the AND/OR semantic counter-intuitive to an actual AND/OR expression in Boolean algebra. However, to be true to the original definition of the fault tree gate, the Boolean expression of AND and OR gates acts upon components' fault state.

\subsubsection{AND Model}
For every state combination of the parent nodes,  we define $\operatorname{Gate}=F$ if all parent nodes are observed to be in state $F$. Hence, the conditional probability distribution for the $\operatorname{Gate}$ node has the following short-hand definition:  
\begin{equation}
	\begin{split}
		&\quad P(\operatorname{Gate} = T | \forall C \in pa(\operatorname{Gate}): C = F ) =  0\\
		&\quad P(\operatorname{Gate} = F | \forall C \in pa(\operatorname{Gate}): C = F ) =  1\\
		&\quad P(\operatorname{Gate} = T | \exists C \in pa(\operatorname{Gate}): C = T ) = 1 \\
		&\quad P(\operatorname{Gate} = F | \exists C \in pa(\operatorname{Gate}): C = T ) = 0 \\
	\end{split}
	\label{eq:and}
\end{equation}

\subsubsection{OR Model}
For every state combination of the parent nodes, we will observe $\operatorname{Gate}=F$ if at least one parent node is in state $F$. 
\begin{equation}
	\begin{split}
		&\quad P(\operatorname{Gate} = T | \forall C \in pa(\operatorname{Gate}): C = T ) = 1\\
		&\quad P(\operatorname{Gate} = F | \forall C \in pa(\operatorname{Gate}): C = T ) = 0\\
		&\quad P(\operatorname{Gate} = T | \exists C \in pa(\operatorname{Gate}): C = F ) = 0 \\
		&\quad P(\operatorname{Gate} = F | \exists C \in pa(\operatorname{Gate}): C = F ) = 1 
	\end{split}
	\label{eq:or}
\end{equation}

\subsubsection{k-out-of-n Model}
For example, Figure~\ref{fig:simplebn} (right side) shows an instance of the k-out-of-n model for a two-out-of-three voting gate. 
The k-out-of-n voting gate triggers a fault event when $k$ or more input events are in a faulty state. Hence, the CPT of the $\operatorname{Gate}$ node has to count how many parent nodes are in the state $F$. This is done for each column. We set the probability to 1 for state $T$ if less than $k$ parent nodes are in the state $F$, or set the probability of $F$ to 1 if $k$ or more parent nodes are in the state $F$. Formally, the conditional probability distribution of the k-out-of-n model has the following definition:
\begin{gather}
	\forall c_1, \dots, c_n \in \{F,T\}^n\nonumber \\ 
	P(\operatorname{Gate} = F| c_1, \dots, c_n) = \begin{cases} 
		1 \quad  \sum_{i=1}^{n}\mathbf {1} _{F}(c_i) \geq k \\ \nonumber
		0  \quad \text{otherwise} \nonumber 
	\end{cases}\\
	P(\operatorname{Gate} = T| c_1, \dots, c_n) = 1-P(\operatorname{Gate} = F| c_1, \dots, c_n) 
	\label{eq:kn}
\end{gather}
where $\mathbf {1} _{F}(x)$ is an indicator function such that
\begin{equation*}
	\mathbf {1} _{F}(x):={\begin{cases}1&{\text{if }} x = F,\\0&{\text{otherwise.}}\end{cases}}
\end{equation*}

\subsection{Transformation Overview}

Algorithm~\ref{alg:appmodel} introduces the pseudo code to build the Bayesian network model based on the high-level service description. Here, the notion $ (x,y,z) \leftarrow S$ means that a  structure, say $S$, provides its elements $x$, $y$, and $z$ to the outer scope, which is known as pattern matching in the context of functional programming.  First we set up a an empty Bayesian network with the node set $X$ and edge set $E$. Afterward, we add our first node $S$, which is a binary random variable representing the availability of the service. At the end of the procedure, one can then infer the fault probability, or availability, of  the service by computing the marginalization $P(\mathcal{S}=F)$, or $P(\mathcal{S}=T)$ respectively. The definition of the conditional probability distribution of  $S$ follows in the procedures in line 7 or 9. 

For any given service model $S$, we build the Bayesian network availability model of the fault dependency graph with the method \textsc{CreateFaultGraph} in line 5, in order to account for cascading and common cause failures, and then include the concrete service type according to $c$.
If $c$ is true, we include the replicated service model with the method  \textsc{ReplicatedService} in line 7, otherwise the procedure builds the redundant service model in line 9.  
The remainder of this section will introduce each of the three sub-procedures in detail. 

\begin{algorithm}
	\caption{Generating the service model}
	\label{alg:appmodel}
	\begin{algorithmic}[1]
		\Procedure{CreateServiceModel}{$S$}
		\State $(\bm{C},Q,G_{\text{FD}},G_{\text{NET}},D,P,\mathcal{G},c)\leftarrow S$ \label{alg:appmodel:10}
		\State $BN = (X,E)$ with $X=\{\}$ and $E=\{\}$  
		\State $X = X \cup \mathcal{S}$ \label{alg:appmodel:20}
		\State \textsc{CreateFaultGraph}($BN$,$G_{\text{FD}}$,$D$,$P$)
		\If{c}
		\State  \textsc{ReplicatedService($BN$,$Q$,$G_{\text{NET}}$,$\mathcal{G}$,$D$)}
		\Else
		\State  \textsc{RedundantService($BN$,$Q$,$G_{\text{NET}}$,$\mathcal{G}$,$D$)}
		\EndIf \label{alg:appmodel:30}
		\State \Return $BN$
		\EndProcedure
	\end{algorithmic} 
\end{algorithm}

\subsection{Fault Dependency Graph}

Given a system model $S$, the first step in the translation procedure is to build the Bayesian network representation of the fault dependency graph. Perhaps it is not apparent why the fault dependency graph forms the beginning. However, due to the cause-effect semantics of Bayesian networks, it is essential to start with root causes first and then successively attach the effects, which themselves are failure causes for other components. Hence, infrastructure failures form the initial causes of failures.

\subsubsection{Failure Model of a Component} 

A component  $C\in \bm{C}$ fails either because of an intrinsic failure or because of an external fault caused by its parent components. First, we define the general Bayesian network structure of a single component. This structure will then be used as a building block for the upcoming  Bayesian network representation of the fault dependency graph.

\begin{figure}[ht]
	\centering
	\includegraphics[scale=0.15]{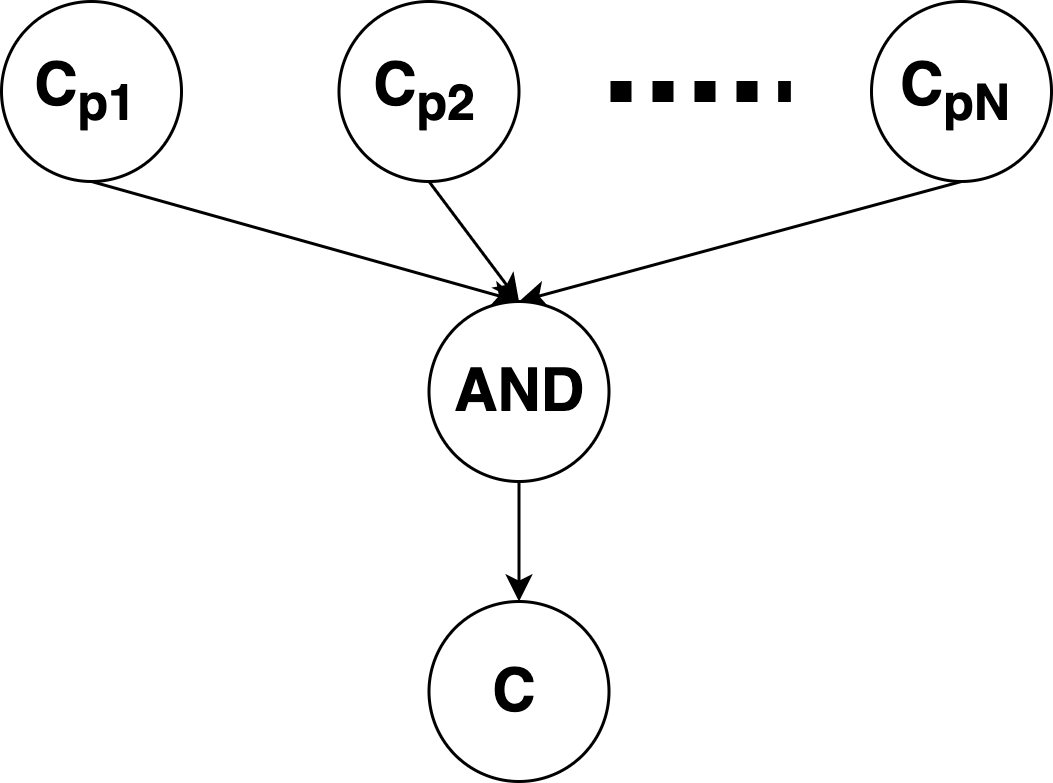}
	\caption{AND fault relation between infrastructure components.}
	\label{fig:and_bn_infrastructure}
\end{figure}

First, the procedure creates a binary random variable for every component in $\mathcal{C}$ with the states $\{F, T\}$, where each variable defines the probability of observing the eponymous component as faulty or available. 
The procedure applies to each component $C$ the Bayesian network transformation of $FT(C)$ according to~\cite{bobbio01}, where the fault of $C$ is the top event, and  $C$'s parent components are the base events.
For example, Figure~\ref{fig:and_bn_infrastructure} shows the Bayesian network representation of a component $C$ that expresses its dependability to its parent components $C_{p_1}$ to $C_{p_N}$ as a fault tree with one AND gate. Hence, the CPT uses the previously introduced AND model from Equation~\ref{eq:and}. A component  $C$ can also fail by its intrinsic fault with probability $q$, which is part of $C$'s CPT definition. The conditional probability distribution of $C$ represents a \textit{noisy}-AND model. Hence, the CPT of $C$ from Figure~\ref{fig:and_bn_infrastructure} has the following definition.
\begin{equation}
	\begin{split}
		&\quad P(C = T | \bm{AND} = F ) = 0 \\
		&\quad P(C= F | \bm{AND}  = F ) = 1 \\
		&\quad P(C= T | \bm{AND} = T ) = 1- q\\
		&\quad P(C= F | \bm{AND} = T ) = q\\
	\end{split}
	\label{eq:faultsemantic}
\end{equation}

\subsubsection{Translating the Fault Dependency Graph} 

Algorithm~\ref{alg:stepone} repeats the approach mentioned above for each component
It transforms a given fault dependency graph  $G_{\text{FD}}$ into a Bayesian network.
First, the procedure creates a node for every component (line~\ref{st3}). Then, it creates their corresponding Bayesian network fault tree representation defined in $FT(C)$ (line~\ref{st7}), using the building formalism introduced by Bobbio et al. in ~\cite{bobbio01}, and then connecting the parent components as base events to the resulting structure at line~\ref{st12}. Finally, we also connect the node of the component that represents the top event (TE) with the corresponding component node (line~\ref{st7.5}). Afterward, it adds the node representation of the instances to the host nodes according to a predefined deployment $D$ (line~\ref{st14}). 

\begin{algorithm}[ht]
	\caption[Build Infrastructure model.]{Building the Bayesian network infrastructure model.}
	\label{alg:stepone}
	\begin{algorithmic}[1]
		\Procedure{CreateFaultGraph}{$BN$,$G_{\text{FD}}$,$D$,$P$}
		\State  $ (\bm{C},E_{\text{INF}},FT) \leftarrow G_{\text{FD}}$
		\For{ $C \in \bm{C}$}\label{st3}
		\State  $X = X \cup C$  \Comment Create node $C$ with state $\{F,T\}$
		\EndFor
		\For{ $C \in \bm{C} \setminus \bm{I}$}
		\State $BN_C = FT(C)$  \label{st7}  \Comment{\parbox[t]{.5\linewidth}{Create Bayesian network model of  $FT(C)$ according to~\cite{bobbio01}}}
		\For{ $C_{pj} \in \text{pa}(C)$}
		\State $E = E \cup (C_{pj}, BN_{C,j})$\label{st12} 
		\EndFor
		\State $E = E \cup (TE(BN_C),C)$ \label{st7.5} 
		\State add CPT to $C$ using  $P$ and Eq.~\ref{eq:faultsemantic} \State with $q=P(I_i =F)$
		\EndFor
		\For{$I_i \in I$} \label{st14}
		\State $E = E \cup (D(I_i),I_i)$ 
		\State add CPT to $I_i$ using $P$ and Eq.~\ref{eq:faultsemantic} \State with $q=P(I_i =F)$
		\EndFor
		\State \Return $BN$ 
		\EndProcedure
	\end{algorithmic} 
\end{algorithm}

\begin{figure}[ht]
	\centering
	\includegraphics[width=\linewidth]{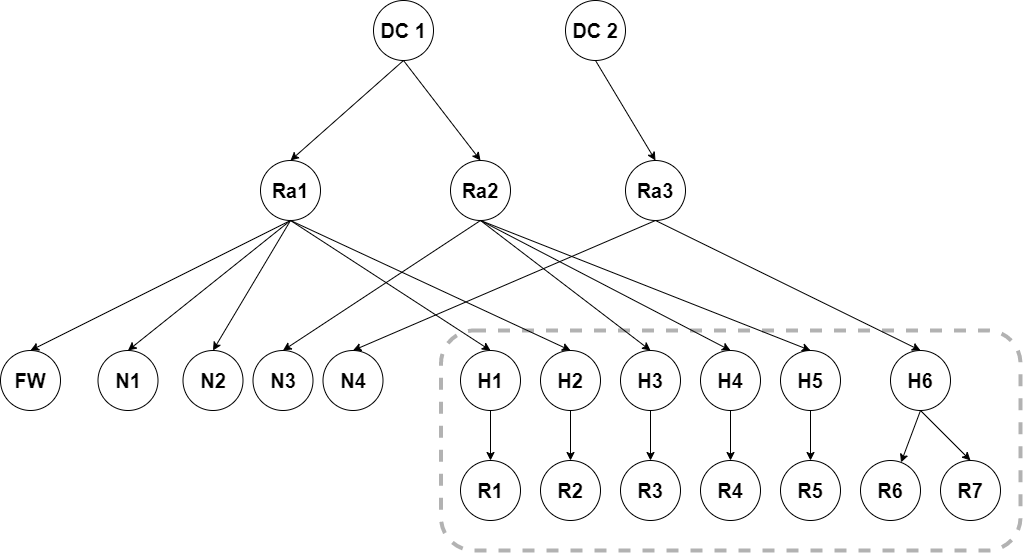}
	\caption{Bayesian network infrastructure model of the data management system example.}
	\label{fig:fdg_bn_infrastructure}
\end{figure}

Applying Algorithm~\ref{alg:stepone} to the example  $S_{\text{Example}}$ leads to the preliminary Bayesian network shown in Figure~\ref{fig:fdg_bn_infrastructure}. 
Here, without loss of generality and for the sake of readability, the AND fault relation between all infrastructure components can be simply combined to one node with the noisy AND model of the component.
With this simplification, the Bayesian network corresponds in its shape to the fault dependency graph, as illustrated in Figure~\ref{fig:infrastructure}. Moreover, to visually assist the translation procedure, the nodes in  Figure~\ref{fig:fdg_bn_infrastructure} are rearranged. All network components are on the left, and all hosts with their processes are on the right side (gray dashed box). 

\subsubsection{Channel Model} 

In order to model service reachability in the presence of network partitioning failures, we need to discuss how to model 
 the probability of observing communication failures with Bayesian networks. Instances and client applications communicate over channels, which is realized as a route along the network graph.
The goal of a channel is to assess the accessibility between two instances in the presence of possible network faults. From an availability perspective,  when a route fails, because some network component had failed along the route, then  a channel can be established along a different if one still exists. Therefore, a channel is considered unavailable, when all potential routes have failed. A channel subsumes the fault probability of observing  all routes  between the two endpoints as interrupted. 

\begin{figure}[h]
	\centering
	\includegraphics[scale=0.15]{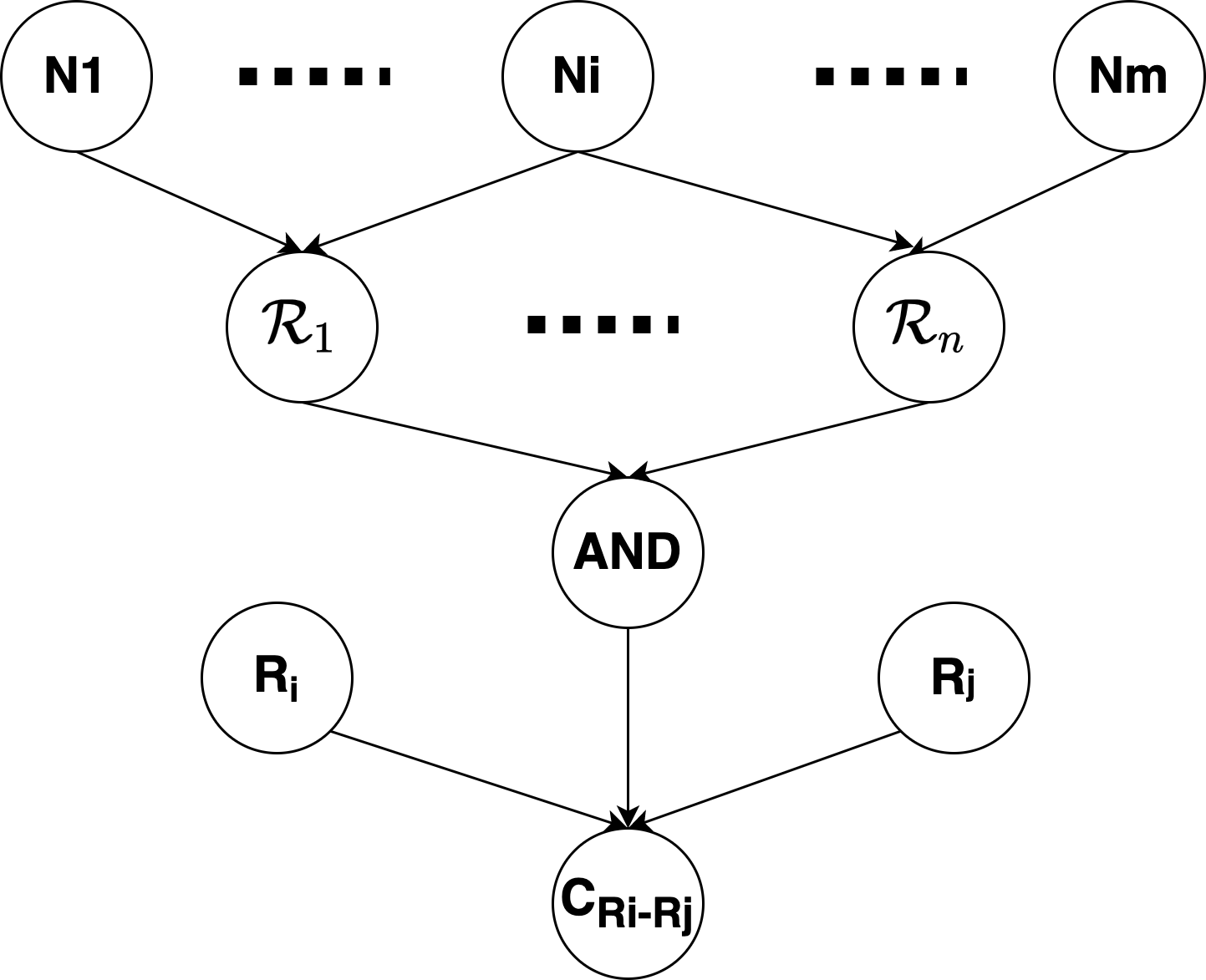}
	\caption{Bayesian network representation of a single communication channel.}
	\label{fig:channel}
\end{figure}

Figure~\ref{fig:channel} shows the Bayesian network structure that contains the 
node $C_{I_i-I_j}$ , representing the  probability of a communication failure between two instances  $I_i$ and $I_j$. 
For readability,  this section refers to $C_{I_i-I_j}$  simply as a \textit{channel node}.
A channel node is conditionally dependent on three nodes: an AND node and two nodes for the endpoints of the channel. The AND node represents the failure probability that no route exists, whereas the endpoint nodes represent the failure probability of the corresponding instances. The CPT of the channel node entails an OR model, defining the probability of observing a channel failure when one of the endpoints fails, or no working route exists. 

The nodes that define the failure of the endpoints, i.e., $I_i$ and $I_j$, are the node representations of the service instances. However, they could also represent different failure causes that indirectly affect the channel, which could be a client application, e.g., the host of the client, or a common endpoint of a second channel. The latter is essential for the replicated service model, to model inner-replica communication.

Finally,  nodes $R_1$ to $R_n$ define the failure probabilities of routes. These route nodes use an OR model for their CPTs and are conditionally dependent on the network components $N_1$ to $N_m$ that are part of the corresponding route in the network graph.  
This model also considers correlating route failures when a route shares the same network components. For example, if $N_i$ fails, route  $R_1$ are $R_n$ are interrupted.  The same applies when multiple channels  share the same routes, respectively. 

Algorithm~\ref{alg:channels} formalizes the construction of a channel as a procedure. Necessary inputs are source $C_{src}$  and destination $C_{dst}$ component and a pair of Bayesian network nodes $X_{src}$ and $X_{dst}$, which represent the failure causes of the channel's endpoints.  
As discussed briefly, the model distinguishes between the components for which it computes the routes and the parent nodes that provide the failure causes at the channel's endpoints.  
The node $AND_{src-dst}$ indicates that the AND node belongs to the channel $C_{src-dst}$, in order to distinguish the AND nodes between multiple channels. First, the procedure computes all routers in the network graph at line~\ref{alg:lin:ch:2}. Afterward, line~\ref{alg:lin:ch:3} to \ref{alg:lin:ch:7} initializes the channel nodes with its parent nodes. Line~\ref{alg:lin:ch:8} iterates over the list of routes and determines if the route has existed as a node in the Bayesian network graph or not. If yes, then the corresponding route node is directly added to the channel as shown in line~\ref{alg:lin:ch:16}. If not, the procedure creates the new route node and connects it with its corresponding network components (lines \ref{alg:lin:ch:10} to \ref{alg:lin:ch:12}). The remainder of the procedure finalizes the  CPT of the channel node and returns it as a reference. 

\begin{algorithm}
	\caption[Create Channels]{Routine to create Bayesian network sub-graph for channels.}
	\label{alg:channels}
	\begin{algorithmic}[1]
		\Procedure{CreateChannel}{( $BN$,  $G_{\text{NET}}$, $C_{src}\!\in\!C_{\text{NET}}$, $C_{dst} \in C_{\text{NET}}$, $X_{src} \in X$, $X_{dst} \in X$)}
		\State $(X, E) \leftarrow BN$
		\State \textit{routes} := compute all paths from $C_{src}$ to $C_{dst}$ in $G_{\text{NET}}$\label{alg:lin:ch:2}
		\State  $X = X \cup C_{src-dst}$ \label{alg:lin:ch:3} 
		\State $X = X \cup AND_{src-dst}$
		\State  $E = E \cup (AND_{src-dst}, C_{src-dst} )$ 
		\State $E = E \cup (X_{src}, C_{src-dst} )$
		\State $E = E \cup (X_{dst}, C_{src-dst} )$ \label{alg:lin:ch:7}
		\For{$\mathcal{R}$ in \textit{routes}} \label{alg:lin:ch:8}
		\If {$\mathcal{R} \notin X$}  \label{alg:lin:ch:10} 
		\State $X = X \cup \mathcal{R}$ 
		\For{$C \in \mathcal{R}.components$}
		\State  $E = E \cup (C, \mathcal{R})$ \label{alg:lin:ch:12}
		\EndFor
		\State add OR model to CPT of $\mathcal{R}$ 
		\EndIf
		\State $E = E \cup (\mathcal{R}, AND_{src-dst})$  \label{alg:lin:ch:16}
		\EndFor
		\State add OR model to CPT  of $C_{src-dst}$ 
		\State add AND model to CPT of $AND_{src-dst}$ 
		\State \Return  $C_{src-dst}$
		\EndProcedure
	\end{algorithmic} 
\end{algorithm}

Without a doubt, the number of routes can get intractably large. In this case, one might resort to simplifying the network graph. That can be done either by aggregating multiple network components, or by considering a limited number of routes --  or both. However, while this simplification increases performance, it comes to the expense of model fidelity. 

\subsection{Redundant Service Model}
Given the channel model, we can build the model of a redundant service first.
Successful communication exists when clients can access sufficient working instances directly. 
In real-life, a client application will most likely try to connect to one instance, whereas the Bayesian network represents the probability of connecting to any of those instances. Due to the high user-load assumption, we need to account for the likelihood of observing sufficient working instances, even if we need one instance to handle the request.

\begin{algorithm}
	\caption{Implementation of the redundant service model.}
	\label{alg:stepthree}
	\begin{algorithmic}[1]
		\Procedure{RedundantService}{$BN$,$Q$,$G_{\text{NET}}$,$\mathcal{G}$,$D$}
		\State $(X, E) \leftarrow BN$
		\State $m = |\mathcal{G}|$
		\For{$i \in [1,m]$}
		\State 	$X = X \cup K_i$ 
		\EndFor
		\For{$G_i \in \mathcal{G}$}
		\For {$I_i \in \bm{I}$}
		\State	$C_{G_i-I_i}\!:=$\textsc{CreateChannel}($BN$,$G_{\text{NET}}$,\\ \quad\quad\quad\quad\quad\quad\quad\quad \quad\quad\quad\quad\quad\quad\quad $G_i$,$D(I_i)$,$G_i$,$I_i$)\label{alg:lin:onst:5}
		\State $E = E \cup (C_{G_i-I_i}, K_i )$ 
		\EndFor
		\State  $E = E \cup (K_i, S)$ 
		\State  add CPT of $K_i$ according to $Q$. 
		\EndFor
		\State add AND model to CPT of $S$ 
		\EndProcedure
	\end{algorithmic} 
\end{algorithm}

Algorithm~\ref{alg:stepthree} describes how to extend the previously created Bayesian network model of the infrastructure with the redundant service model. 
We stated in the system model, that a client application can access a service through one or more dedicated network component that act as entry points, i.e. gateways, in the network. Therefore, we introduce a new set of binary random variables $K = \{K_i\}_{i=1}^m$, with $m=|\mathcal{G}|$, which represents the probability of accessing sufficient instances through the i-th entry point defined in $\mathcal{G}$.

\begin{figure}[ht]
	\centering
	\includegraphics[width=\linewidth]{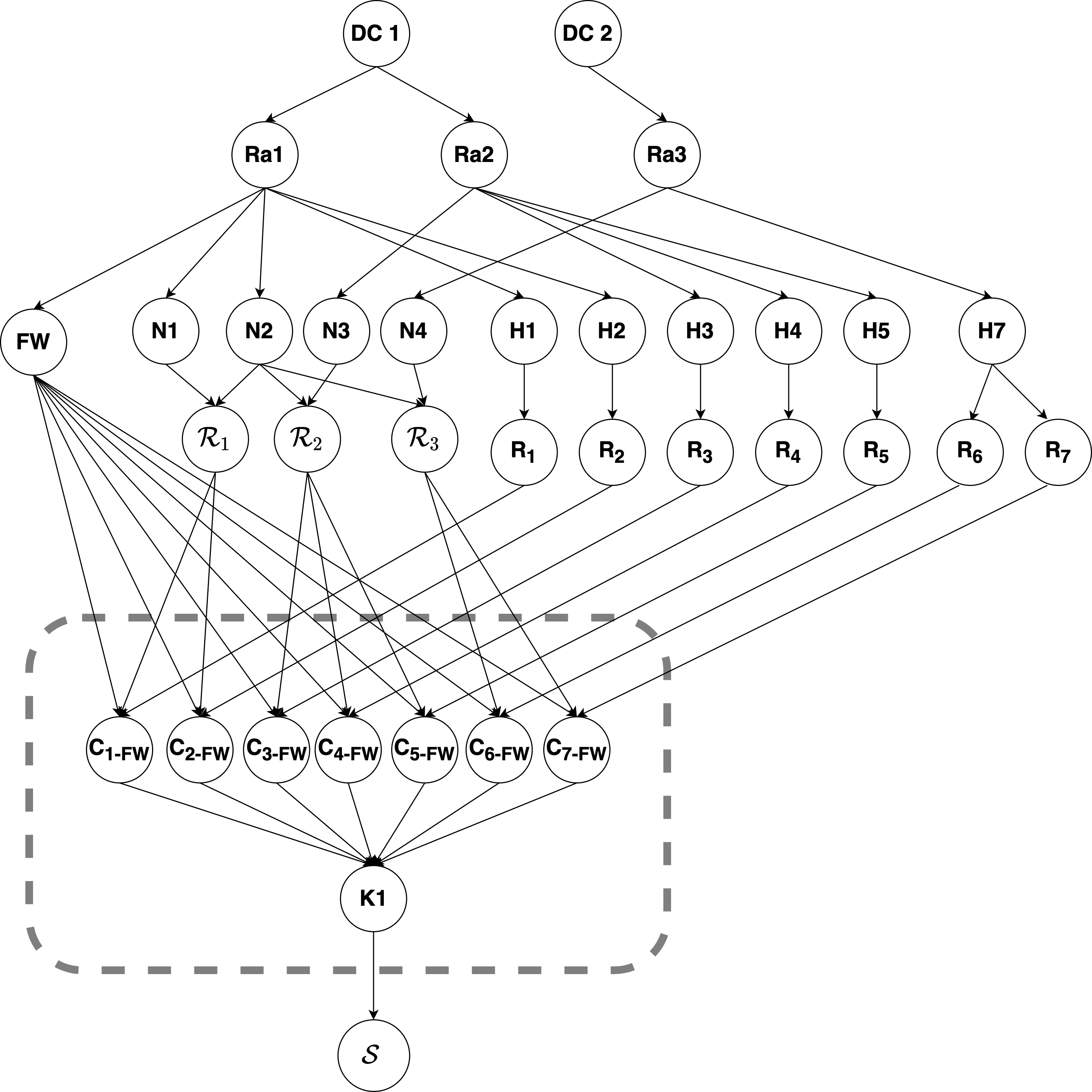}
	\caption{The Bayesian network of a redundant service example.}
	\label{fig:readone}
\end{figure}

At line~\ref{alg:lin:onst:5}, the procedure creates the channel nodes for each entry point in the set $\mathcal{G}$ to each instance. The channel creation procedure takes as input the network component that acts as an entry point, the host of the instances as defined by their deployments, and the two nodes that represent the failure of the channel's endpoints. 
In a follow up step (line 11), all channel nodes that are related to the i-th entry point component connect to one node $K_i$, which implements the reachability requirement of accessing sufficient instances form that  entry point as part of its CPT. 
For example, if one instance is sufficient for a working service, then each  $K_i$ would implement an AND model at line 14, representing the fault probability that the i-th client cannot communicate with any instance at all. 
A detailed discussion on how to integrate general requirements for $K_i$ at line 14 can be found at the end of this section. 

Finally, Algorithm~\ref{alg:stepthree} finishes by introducing the final service node $\mathcal{S}$. This node accounts for the probability that no client at any entry point has sufficient working channels to communicate with the instances. Hence, we can compute the probability of a single service failure as the marginal $P(\mathcal{S}=F)$ or its availability $P(\mathcal{S}=T)$ using Bayesian inference.

For instance, Figure~\ref{fig:readone} shows the Bayesian network model  of the example service $\mathcal{S}$ from Section~\ref{sec:model}, assuming a redundant service.
In this example, all clients communicate with the instances via the firewall (represented by node \texttt{FW}). There are three routes $\mathcal{R}_1$ to $\mathcal{R}_3$, which are shared by all seven channels, emphasized by the dashed box. Each channel is connected to the firewall node, representing the client. Since there is only one entry point, the set $K=\{K_1\}$ contains one node. For example, assuming the service can tolerate three instance failures, node $K_1$ implements a four-out-of-seven model (see Equation~\ref{eq:kn}). 

\subsection{Replicated Service Model}
\label{sec:2step}
For replicated services, we said that clients first send their request to one instance, which then communicates with the remaining instances. This communication pattern subsumes and implements the likelihood of accessing at least one instance that can communicate with sufficient remaining instances. Hence, we will show how to use this communication pattern to encode all possible states in which the instances, or cannot, reach the desired number of votes, e.g, quorum size, as defined by the fault tolerance model in $Q$. Consequently, the final Bayesian network will encode the probability of observing the service in a working state, giving potential infrastructure and communication faults. 

Algorithm~\ref{alg:stepfour}  begins first by modeling the communication channels between instances. It introduces
again the set of  binary random variables $K = \{K_i\}_{i=1}^n$ where $n = |R|$, which represent the failure probability of communicating with an insufficient number of instances when the i-th instance initiates the replication protocol.
Hence, every $K_i$  is a child node of $n-1$ channel nodes (line~\ref{sf4} and~\ref{sf5}), since the fault probability of instance $R_i$ is already part of one of the endpoints of the channels. 
Next, the procedure builds  a channel node for every  entry point $G_i$ to every instance $R_i$ by using  $K_i$ as failure cause(line~\ref{sf10}).  Instead of directly addressing the failure probability of an instance, the model uses $K_i$ to represent the instance $R_i$. In the case of a network partitioning, $K_i$ would contain the probability that $R_i$ can still access sufficient processes in its partition. 

Finally,  node $\mathcal{S}$ accounts for the failure probability that no client can access the service through any gateway(line~\ref{sf14}). Hence, one can now infer the fault probability, or availability, of  the service by computing the marginalization $P(\mathcal{S}=F)$, or $P(\mathcal{S}=T)$ respectively.

\begin{algorithm}
	\caption{Implementation of the replicated service model.}
	\label{alg:stepfour}
	\begin{algorithmic}[1]
		\Procedure{ReplicatedService}{$BN$,$Q$,$G_{\text{NET}}$,$\mathcal{T}$,$D$}
		\State $(X, E) \leftarrow BN$
		\State $X = X \cup \mathcal{S}$  \Comment Create  service node $\mathcal{S}$ 
		\State $n = |R|$ 
		\For{$i \in [1,n]$}
		\State 	$X = X \cup K_i$ \label{sf1}
		\EndFor
		
		\For{ $(R_i,R_j)$ in $R \times R$} 
		\If{$C_{R_i-R_j} \notin X$ and $C_{R_j-R_i} \notin X$} 

		\State $C_{R_i-R_j}$ := \textsc{CreateChannel}($BN$,$G_{\text{NET}}$,\\ \quad\quad\quad\quad\quad\quad\quad\quad \quad\quad\quad\quad\quad$D(R_i)$,$D(R_j)$,$R_i$,$R_j$)
		\State $E = E \cup (C_{R_i-R_j}, K_i )$\label{sf4} 
		\State $E = E \cup (C_{R_i-R_j}, K_j )$\label{sf5}   
		\EndIf
		\EndFor
		\State  add CPT for all $K_i \in K$ according to $Q$. 
		\For{ $G_i \in \mathcal{G}$} 	\Comment Second Step
		\State $X = X \cup G_i$
		\For{j=1; j < n; j++}
		\State $C_{G_i-P_j}$ := \textsc{CreateChannel}($BN$,$G_{\text{NET}}$, \\ \quad\quad\quad\quad\quad\quad\quad\quad \quad\quad\quad\quad\quad $G_i$,$D(R_j)$,$G_i$,$K_j$) \label{sf10} 
		\State  $E = E \cup (C_{G_i-R_i}, \mathcal{S} )$  
		\EndFor
		\EndFor
		\State add AND model to CPT of $\mathcal{S}$  \label{sf14} 
		\EndProcedure
	\end{algorithmic} 
\end{algorithm}

\begin{figure}[h]
	\centering
	\includegraphics[width=\linewidth]{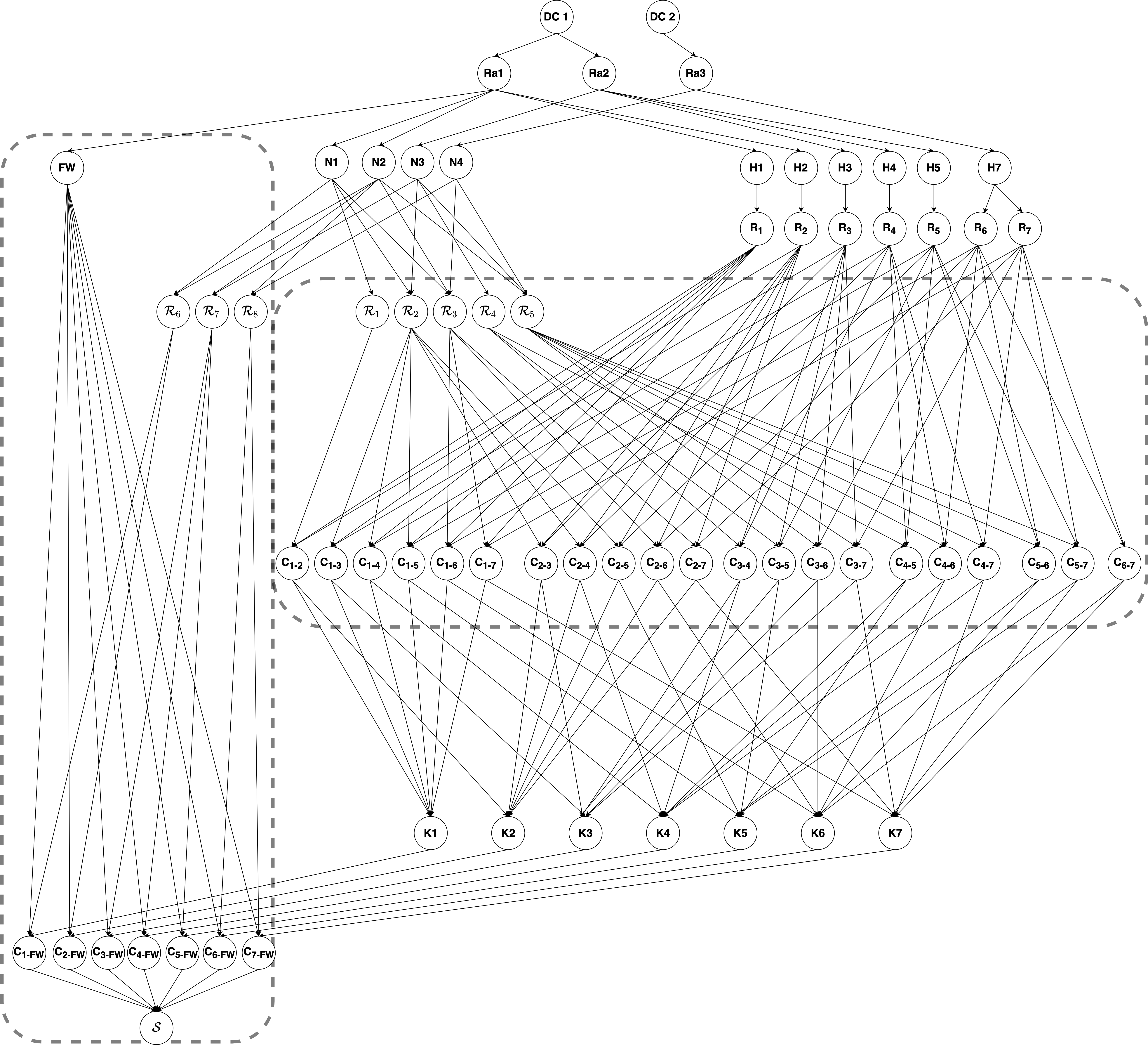}
	\caption{The Bayesian network of the indirect communication pattern for the database example.}
	\label{fig:replication}
\end{figure}

For example, Figure~\ref{fig:replication} shows the Bayesian network of the database service example, based on the assumption that client applications access the service via the firewall. The left box shows the channel nodes representing the fault probabilities for the communication between clients and service instances. The right box shows the channels of each instance to every other instance. A node $K_i$ has as parent nodes the channel nodes of the i-th instance. Hence, to implement the majority set requirement, one can use a three-out-of-six model for $K_i$ to encode the probability of observing at least three working channels, which implies that the i-th instance is also working.

Next, we discuss in detail how to implement the CPTs of the nodes in $K$ as hinted at line 16. 
\subsubsection{Read-one/write-all}
\label{sec:algos}
Read-one/write-all is a special case in replication since every operation has its particular quorum requirements. We already had a brief introduction on read-one/write-all in the last section. There, we discussed how to implement the service requirements for read  $Q_{ro} = 2^R/\emptyset$, and for write quorums $Q_{wa} = \{R\}$.
Consequently, each operation needs its own Bayesian network model to assess its availability individually.
Read-one can be modeled by using the redundant service model. Hence, the model uses an AND model for all CPTs of the nodes in $K$ to account for the fault probability that no channel works. 
In contrast, for write-all, it depends on the system design. One can use either the redundant or th replicated service model. Both models use the OR model for the CPTs of nodes in $K$, accounting for the fault probability that there is at least one channel faulty to an instance.

\subsubsection{K-out-of-N Voting}

In voting-based replication, instances have one vote to decide on an incoming operation request. The system is available when it can reach k-out-of-n votes for some request, e.g., majority sets require $k=\lfloor\frac{n}{2}\rfloor+1$ votes.
For replicated services that use the indirect communication pattern, the i-th replica is part of the voting process, where it must acquire at least $k-1$ votes from the remaining $n-1$ replicas to consider the service as available. Thus, the CPT of  $K_i$  implements an  $n-k+1$-out-of-$n-1$ mode as defined in Equation~\ref{eq:kn}l, i.e.,  considering the inverse on how many channel failures can be tolerated. 

For redundant services that use the direct communication pattern with $n$ instances, where $k$ instances are sufficient to signify that the service does not fail due to overload, the model implements the CPT of  $K_i$ by using an $(n-k)$-out-of-$n$ model. Thus, the system fails if there are more than $n-k$ channels faulty. 

\subsubsection{Weighted Voting}
In weighted voting, individual replicas can have multiple votes. This forms the general case of the normal voting-based appraoch from above.
To reach a potential quorum, the total number of votes that are available by working instances needs to exceed a given threshold $t$.
As a result, this work extends the k-out-of-n model from Equation~\ref{eq:kn} to account for the individual vote counts of the replicas. We use the tuple notation for $Q = (V,t)$, where $V = (v_1,...,v_n)$ are instance votes and $t$ the threshold value.
Given that $K_i$ refers to the i-th instance, the models use $v_j$ to denote the number of votes of the instance at the opposing endpoint of the j-th channel for a given state combination  $c_{i-1}, \dots, c_{i-m} $ of the channel nodes connected at $K_i$.
Here, since the i-th instance initiated the replication protocol, we automatically assume that its votes $v_i$ contribute to the request. Hence we reduce the threshold by its votes.
\begin{gather*}
	\forall c_{i-1}, \dots, c_{i-m}  \in \{F,T\}^m\\
	P(K_i \!=\! T| c_{i-1}, \dots, c_{i-m} ) = \begin{cases} 
		1 \quad   \sum_{j=1}^{m}\mathbf {1} _{T}(c_{i-j})v_j \!\geq\! t \!-\! v_i \\
		0  \quad \text{otherwise} 
	\end{cases}\\
	P(K_i\! =\! F| c_{i-1}, \dots, c_{i-m} ) = 1 - P(K_i \!=\! T| c_{i-1}, \dots, c_{i-m} )
\end{gather*}

For every state combination  $c_{i-1}, \dots, c_{i-m}$, the model builds the weighted sum of those channels that are available and checks if the result is above the threshold. 

\subsection{Scalability}

Bayesian networks are subject to an exponential growth of memory with regard to their CPTs~\cite{koller09}. The CPT of a node has to implement a conditional probability distribution for each state combination of its parent nodes. If the parent nodes are binary, then the number of CPT entries is $O(2^n)$. Hence, all CPTs of $K$ will exhibit an exponential memory growth in the number of instances. We have a similar situation for nodes that represent the availability of routes. Those nodes implement an OR model, which can have multiple network components that represent a route. Assuming a CPT entry is just several bytes large, it is not hard to see that a node with 30 parent nodes will have a CPT with several gigabytes of memory. Therefore, this  Bayesian network approach is suitable only for services with up to 30 instances and short network routes; afterward, the memory becomes the limiting factor.

However, this problem can be mitigated for the AND/OR, and k-out-of-n model. Heckerman~\cite{heckerman93} provides an equivalent AND/OR model that reduces the space complexity to linear, while Bibartiu et al.~\cite{bibartiu21} provide an equivalent (scalable) k-out-of-n model with polynomial complexity. Having these scalable models, we can substitute the existing AND/OR, and k-out-of-n models in the Bayesian network model with their scalable counterparts. Hence, we can overcome the memory limitations for redundant services and voting-based replication models for large services.

\section{Evaluation}
\label{sec:eval}

\begin{figure*}[th]
	\centering
	\begin{tikzpicture}
		\begin{groupplot}[%
			group style={group size=2 by 1},
			]
			\nextgroupplot[xmin=0,xmax=300,ymajorgrids=true, xmajorgrids=true,title={Redundant Service},
			legend style={at={($(0,0)+(1cm,1cm)$)},legend columns=2,fill=none,draw=black,anchor=center,align=center},
			legend to name=fred,
			ylabel={Availabiltiy}
			] 
			\coordinate (c1) at (rel axis cs:0,1); 
			\addplot table[x=n,y=sc_approx,col sep=comma] {eval/service/simple/redundant/final_availability.csv};
			\addlegendentry{Small Infrastructure  with Approx. Inference}
			\addplot table[x=n,y=sc_approx,col sep=comma] {eval/service/large/redundant/final_availability.csv};
			\addlegendentry{Large Infrastructure with Approx. Inference}
			\addplot table[x=n,y=sc_exact,col sep=comma] {eval/service/simple/redundant/final_availability.csv};
			\addlegendentry{Small Infrastructure  with Exact Inference}
			\addplot table[x=n,y=sc_exact,col sep=comma] {eval/service/large/redundant/final_availability.csv};
			\addlegendentry{Large Infrastructure  with Exact Inference}
			\nextgroupplot[xmin=0,xmax=300,ymajorgrids=true, xmajorgrids=true,title={Replicated Service}]
			\coordinate (c2) at (rel axis cs:1,1); 
			\addplot table[x=n,y=sc_approx,col sep=comma] {eval/service/simple/replicated/final_availability.csv};
			\addplot table[x=n,y=sc_approx,col sep=comma] {eval/service/large/replicated/final_availability.csv};
			\addplot table[x=n,y=sc_exact,col sep=comma] {eval/service/simple/replicated/final_availability.csv};
			\addplot table[x=n,y=sc_exact,col sep=comma] {eval/service/large/replicated/final_availability.csv};
		\end{groupplot}
		\coordinate (c3) at ($(c1)!.5!(c2)$);
		\node[below,yshift=-0.8cm,xshift=-0.5cm] at (c3 |- current bounding box.south)
		{\pgfplotslegendfromname{fred}};
		\node[below,yshift=2.2cm] at (c3 |- current bounding box.south) {\# Instances};
	\end{tikzpicture}
	\caption{The availability results of a service for increasing the number of instances, using approximate and exact inference.}
	\label{fig:availability}
\end{figure*}
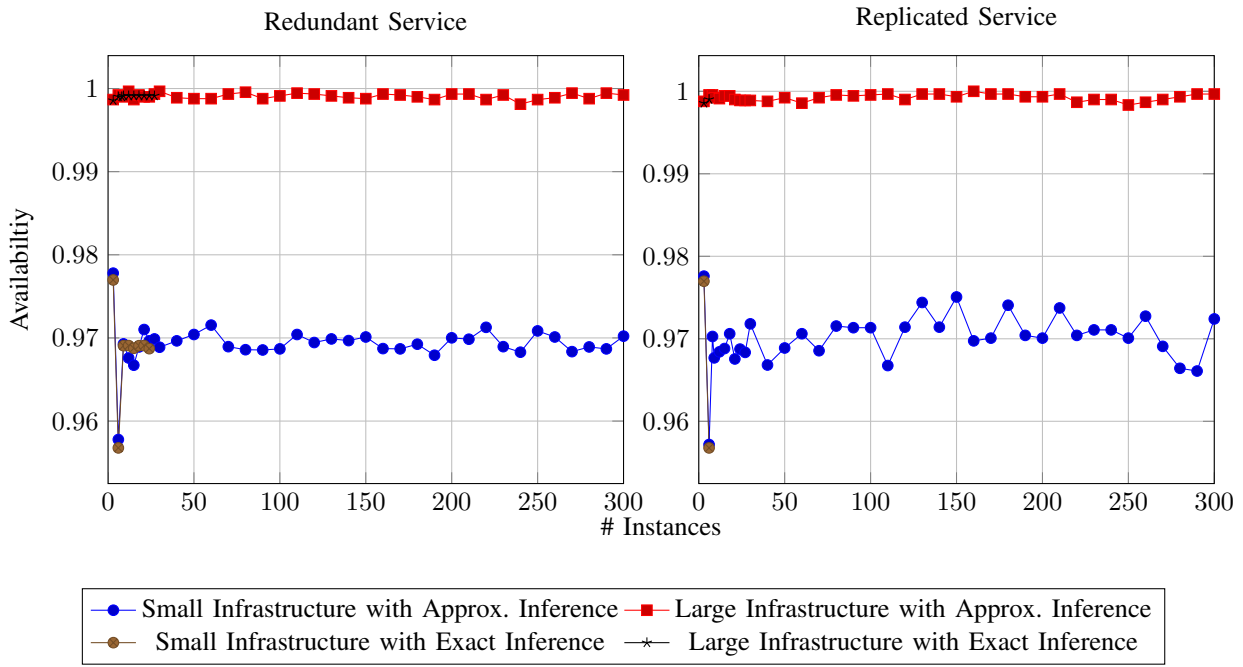
This section provides an in-depth analysis of the performance and modeling feasibility of the presented Bayesian network availability model. The evaluation will analyze the availability, build, and inference performance for redundant and replicated services for an increasing number of instances. All experiments were performed on a 64-bit machine with 64  Intel(R) Xeon(R) CPU E7-4850 v4 at 2.10GHz and 1 TB of main memory, running Arch Linux  5.13.12 with GCC 11.1.0, Python 3.9.6, and with pgmpy 0.1.7 (the Bayesian network modeling package) and Numpy 1.20.3. Bayesian network inference is performed with approximate and exact inference whenever possible. For approximate inference, we use the forward sampling method, and for exact inference, we used the Lauritzen-Spiegelhalter Algorithm method~\cite{lauritzen88} from the gRain 1.3.2 package ~\cite{gRain,gRainb}. Furthermore, we used in all experiments the scalable Bayesian network representations for AND/OR and voting gates by Heckerman~\cite{heckerman93} and Bibartiu et al.~\cite{bibartiu21}. The implementation of the algorithms and evaluation methods for the presented Bayesian network model are available as open source\footnote{https://github.com/openclams/bn-availability-model}. 

Moreover, all experiments will consider two different data center infrastructures. The first infrastructure corresponds to the example used in Section~\ref{sec:model}, which consists of  19 components. The evaluation will refer to this example as the \textit{small} infrastructure. Consequently, the second infrastructure will be called the \textit{large} infrastructure. The large infrastructure consists of three data centers with 40 hosts each, using a random topology of 20 network components to connect hosts and data centers.
Moreover, each data center has  100  additional infrastructure components that influence the hosts and the network components. The large infrastructure has in total 440 components. All components in the large infrastructure have an availability value sampled from a beta distribution with $C \sim  \operatorname{Beta}(10000,1)$, resulting in an average downtime of 1 hour during a mission time of 10,000 hours. Without loss of generality, we will require that the majority of instances are needed for both service types to be considered available. Other k-out-of-n schemes are also possible, but a different $k$ changes only the content of the corresponding nodes and not the structure of the Bayesian network. 

The plot in Figure~\ref{fig:availability} shows the expected availability for both service types for an increasing number of instances, using the small and large infrastructure, applying approximate and exact inference. 
Instances were placed in round-robin. We computed the availability for services with up to 300 instances using approximate inference. Exact inference was only possible for up to 27 instances for the redundant service experiments and for up to 6 instances for the replicated service experiments, independently of the infrastructure size. Approximate inference might vary by nature with every execution. So we compared the results of the exact and approximate inference methods by repeating them 40 times to compute their confidence intervals. As a result,  it can be stated with 95\% confidence that there is no significant difference in the inference results between the exact and approximate inference methods here.

The availability results between the redundant and the replicated service are similar. 
The availability decreases up until six instances for the small infrastructure experiments. This is mainly because all instances are placed in the first data center. The follow-up placements also consider the second data center in the small infrastructure for services with seven or more instances. The more instances,  the less common-cause failures are shared. However, adding more instances does not lead to higher availability. The higher the distribution of instances, the higher the risk of communication failures since more network components are involved.  This limits the availability to a point where the influence of the shared infrastructure outweighs the benefits of replication. Even in the large infrastructure example, where we assume a low average downtime per component, the availability does not converge arbitrarily near to 1.

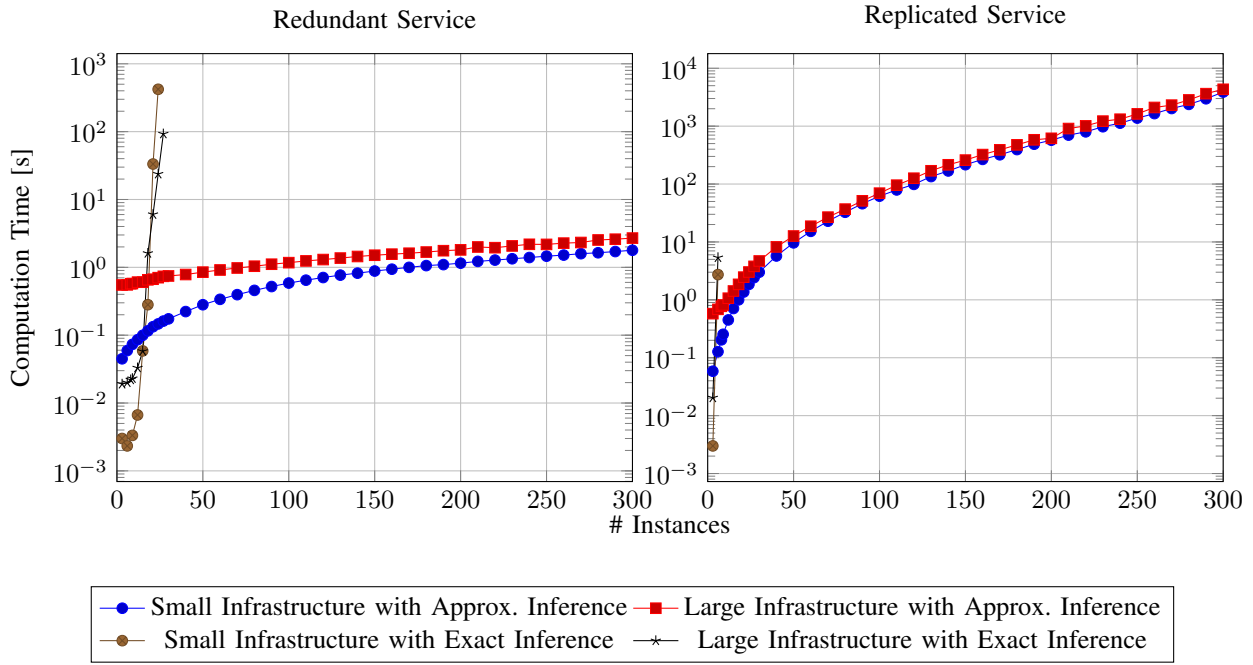
\begin{figure*}[ht]
	\centering
	\begin{tikzpicture}
		\begin{groupplot}[
			group style={group size=2 by 1},
			]
			\nextgroupplot[ymode=log,xmin=0,xmax=300,ymajorgrids=true, xmajorgrids=true,title={Redundant Service},
			legend style={at={($(0,0)+(1cm,1cm)$)},legend columns=2,fill=none,draw=black,anchor=center,align=center},
			legend to name=fred,
			ylabel={Computation Time [s]}
			] 
			\coordinate (c1) at (rel axis cs:0,1);
			\addplot table[x=n,y=sc_approx,col sep=comma] {eval/service/simple/redundant/final_inference_time.csv};
			\addlegendentry{Small Infrastructure  with Approx. Inference}
			\addplot table[x=n,y=sc_approx,col sep=comma] {eval/service/large/redundant/final_inference_time.csv};
			\addlegendentry{Large Infrastructure with Approx. Inference}
			\addplot table[x=n,y=sc_exact,col sep=comma] {eval/service/simple/redundant/final_inference_time.csv};
			\addlegendentry{Small Infrastructure  with Exact Inference}
			\addplot table[x=n,y=sc_exact,col sep=comma] {eval/service/large/redundant/final_inference_time.csv};
			\addlegendentry{Large Infrastructure  with Exact Inference}
			\nextgroupplot[ymode=log,xmin=0,xmax=300,ymajorgrids=true, xmajorgrids=true,title={Replicated Service}]
			\coordinate (c2) at (rel axis cs:1,1); 
			\addplot table[x=n,y=sc_approx,col sep=comma] {eval/service/simple/replicated/final_inference_time.csv};
			\addplot table[x=n,y=sc_approx,col sep=comma] {eval/service/large/replicated/final_inference_time.csv};
			\addplot table[x=n,y=sc_exact,col sep=comma] {eval/service/simple/replicated/final_inference_time.csv};
			\addplot table[x=n,y=sc_exact,col sep=comma] {eval/service/large/replicated/final_inference_time.csv};
		\end{groupplot}
		\coordinate (c3) at ($(c1)!.5!(c2)$);
		\node[below,yshift=-0.8cm,xshift=-0.5cm] at (c3 |- current bounding box.south)
		{\pgfplotslegendfromname{fred}};
		\node[below,yshift=2.2cm] at (c3 |- current bounding box.south) {\# Instances};
	\end{tikzpicture}
	\caption{The  inference time to compute the availability of a service with increasing number of instances for the small and large infrastructure example, using approximate and exact inference.}
	\label{fig:inference}
\end{figure*}

The plot in Figure~\ref{fig:inference}  shows the mean inference time to compute the presented availabilities. Here we can observe the exponential time increase (linear function in a semi-log plot) of the exact inference method, which contrasts the polynomial time increase (log function in a semi-log plot) of the approximate inference method. There are two main observations. First, the inference time between the redundant and replicated services have different polynomial complexities, and second, the inference time converges independently of the infrastructure size. Clearly, due to the twenty-fold increase of components in the large infrastructure compared to the small infrastructure, the former is slower than the latter for small numbers of instances. However, the number of channels nodes increases with the number of instances. Hence, the more instances, the more channel nodes. The number of channel nodes outweighs the number of infrastructure components until they become the influencing factor in the computation. The model of the redundant service has a linear increase of channels, whereas the replicated service has a quadratic increase of channel nodes due to the indirect communication pattern.  

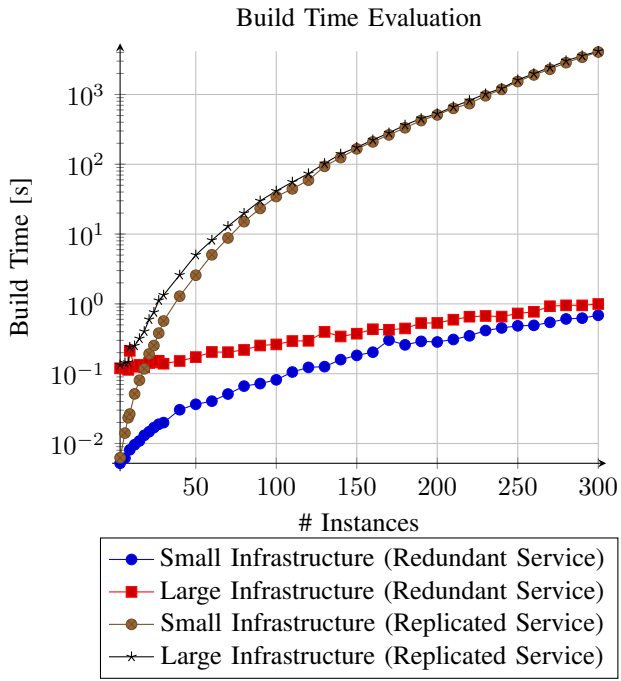
\begin{figure}[ht]
	\centering
	\begin{tikzpicture}
		\begin{semilogyaxis}[%
			width=\linewidth-2.5cm,
			scale only axis,
			axis line style={shorten >=-3pt, shorten <=-3pt}, 
			axis lines=left, 
			legend columns=1,
			legend style={cells={align=left},at={(0.5,-0.18)},anchor=north},  
			ymajorgrids=true, xmajorgrids=true, 
			title={Build Time Evaluation},
			xlabel={\# Instances},
			ylabel={Build Time [s]},
			]
			\addplot table[x=n,y=sc_approx,col sep=comma] {eval/service/simple/redundant/final_build_time.csv};
			\addlegendentry{Small Infrastructure (Redundant Service)}
			\addplot table[x=n,y=sc_approx,col sep=comma] {eval/service/large/redundant/final_build_time.csv};
			\addlegendentry{Large Infrastructure (Redundant Service)}
			\addplot table[x=n,y=sc_approx,col sep=comma] {eval/service/simple/replicated/final_build_time.csv};
			\addlegendentry{Small Infrastructure (Replicated Service)}
			\addplot table[x=n,y=sc_approx,col sep=comma] {eval/service/large/replicated/final_build_time.csv};
			\addlegendentry{Large Infrastructure (Replicated Service)}
		\end{semilogyaxis}
	\end{tikzpicture}
	\caption{Comparing the time to build the Bayesian network model for increasing number of instances.}
	\label{fig:buildtime}
\end{figure}

Finally, Figure~\ref{fig:buildtime} introduces the build time to construct the Bayesian network. Clearly, the build time shows a significant difference between the large and small infrastructure examples for small numbers of instances with $n$ less than 30 w.r.t. service type. However, with increasing numbers instances the time difference diminishes. Afterwards, the sole factor that determines the build time is the service type. For large numbers of instances, the infrastructure has almost not significant influence on the build time anymore. Again the number of channel nodes that grow in proportion to the number of instances outweighs the component nodes of the infrastructure.  

\section{Discussion}
\label{sec:discussion}

The evaluation demonstrated the feasibility of the Bayesian network approach to model large-scale and replicated systems. Build and inference time is within manageable time frames for reliability engineers to make informed decisions on the service. Overall, for small service sizes with three to seven replicas as commonly used for transaction-oriented database systems, the reliability engineer can even use exact inference to assess the availability in order to compute deterministic results. We discussed that the number of channels has the most influence with regard to the build and inference time. Replicated services lead to a quadratic growth of channel nodes in the number of instances. Also, the build procedure needs to compute all possible routes that constitute a channel. 
Finding all possible routes in a graph can become a performance impediment, which is why we suggest considering only a subset of essential routes if performance is of higher priority. The largest model with 300 instances took about one hour to build. But once the model is built, inference can be performed independently often. Even updating individual beliefs of component failures can be done directly to the respective nodes if needed, without rebuilding the whole Bayesian network.

A particular modeling challenge is the potential lack of accurate availability data (failure probabilities). Acquiring accurate failure data is a non-trivial task for rare events, which require a large number of observations to conclude statistical significance. However, this issue can be addressed in several ways. First, many vendors already provide mean time to failure (MTTF) information for their software or hardware components. Secondly, cloud providers host larger numbers of hardware components in their data centers, which are constantly monitored, providing significant amounts of data also for rare events~\cite{peter21}. Thirdly, for yet unobserved failures of highly available components, one can use rare event analysis (an active research area) in conjunction with expert knowledge acquisition to incorporate prior beliefs first and later refine the estimate with observation during mission time. 

Moreover, our model does not consider the effects of long-running requests and the implications of component failures and recoveries during a longer execution time. This would require a dynamic Bayesian network approach~\cite{boudali05,boudali05a} to model the time dimension, bringing new challenges w.r.t. model assumptions, which might require additional implementation details of the particular replication protocol, increasing the model complexity. Therefore, we consider this challenge as future work.

\section{Related Work}
\label{sec:rw}
Modeling complex infrastructures is subject to various areas of reliability engineering~\cite{kim09,kim11,ford10,narayanan17a}. Jammal et al.~\cite{jammal16} provide a hierarchical infrastructure model for cloud services with Petri nets as an evaluation framework. They consider fault propagation within a hierarchical infrastructure model supporting redundant cloud services with a one-out-of-n fault tolerance semantic. However, they do not consider network communication.

Ghosh et al.~\cite{ghosh14}, and Narayanan et al.~\cite{narayanan17a} consider a k-out-of-n redundancy model for their instances; however, their infrastructure model only considers fault-independent compute nodes or data centers, respectively.  

The Palladio Component model~\cite{brosch12,becker09}  provides a holistic modeling approach to evaluate the performance and availability of complex software systems unifying hard- and software into one model. However, the Palladio availability model supports only a one-out-of-n redundancy model and cannot model quorum requirements.

There are several methods to evaluate the availability of a system, among which Bayesian networks have gained large acceptance within the industry and research~\cite{torres98,ye20,cai19,kammouh20}.  

Bobbio et al.~\cite{bobbio99,bobbio01} demonstrated the applicability and superiority of Bayesian networks in modeling and evaluating equivalent fault trees~\cite{ruijters15}.
Moreover, Boudali and Dugan~\cite{boudali05, boudali05a}  showed how to use dynamic Bayesian networks to model dynamic fault trees as well, effectively proving that the Bayesian network formalism is powerful enough to cover all non-state space models.

Bennacer et al.~\cite{bennacer15} use Bayesian networks for network diagnostics by introducing a case-based reasoning inference approach to increase diagnostic performance for large-scale Bayesian network models. While they only focus on network communication, they provide a tailored inference technique for efficient diagnostics of root causes, which can also be combined with our Bayesian network model when diagnostics is of interest.

Pitakrat et al.~\cite{pitakrat17} use Bayesian networks for online failure predictions of microservice applications. The Bayesian network represents the interconnection between the microservice instances and updates the fault probabilities of the services based on the online monitoring of performance metrics. They consider fault propagation between services; however, replication is not considered. 

In summary, a Bayesian network modeling approach, covering a wide range of redundant and replicated services that also includes cascading and correlated faults caused by dependent infrastructure and network communication, was missing.

\section{Conclusion}
\label{sec:conclusion}
This work introduced a Bayesian network availability model for redundant and replicated services. The Bayesian network model unifies the fault aspects defined within a high-level model description of the service. The high-level model consists of three sub-models: a fault dependency graph to express the failure relation between components of the infrastructure and execution environment, a network model to address communication and network partitioning failures, and a  model to define fault-tolerance requirements of the service. We show how to translate the high-level model  into one Bayesian network to compute the expected availability. Finally, evaluations demonstrate the feasibility of the Bayesian network approach to represent and assess the availability of large-scale service with hundreds of fault influences and service instances. 

\section{Acknowledgments}

This work was supported by the Robert Bosch GmbH.

\bibliographystyle{IEEEtranDOI}
\bibliography{main}

\end{document}